\documentclass{SciPost}

\hypersetup{
    colorlinks,
    linkcolor={red!50!black},
    citecolor={blue!50!black},
    urlcolor={blue!80!black}
}

\usepackage[bitstream-charter]{mathdesign}
\usepackage{orcidlink} 

\DeclareSymbolFont{usualmathcal}{OMS}{cmsy}{m}{n}
\DeclareSymbolFontAlphabet{\mathcal}{usualmathcal}

\fancypagestyle{SPstyle}{
\fancyhf{}
\lhead{\colorbox{scipostblue}{\bf \color{white} ~SciPost Physics }}
\rhead{{\bf \color{scipostdeepblue} ~Submission }}

\fancyfoot[C]{\textbf{\thepage}}
}

\usepackage{tabularray}
\usepackage{bm}
\usepackage{slashed}
\usepackage{enumitem}

\newcommand{\mbf}[1]{\mathbf{#1}}
\newcommand{\dd}{\mathrm{d}}
\newcommand{\Tr}{\mathop{\mathrm{Tr}}}
\newcommand{\vi}{{\boldsymbol{i}}}
\newcommand{\vj}{{\boldsymbol{j}}}
\newcommand{\vecr}{{\boldsymbol{r}}}
\newcommand{\vk}{{\boldsymbol{k}}}

\newcommand{\hatbold}[1]{\hat{\mbf{#1}}}
\newcommand{\haty}{{\mbf{\hat{y}}}}
\newcommand{\hatx}{{\mbf{\hat{x}}}}

\newcommand{\X}{{\mathcal{X}}}

\definecolor{colone}{HTML}{fec4de}
\definecolor{coltwo}{HTML}{cbf0f6}
\definecolor{colthree}{HTML}{c7ece0}

\newcommand{\SO}{\mathrm{SO}}
\newcommand{\SU}{\mathrm{SU}}
\newcommand{\U}{\mathrm{U}}
\newcommand{\Sp}{\mathrm{Sp}}

\newcommand{\eqnref}[1]{Eq.~\eqref{#1}}
\newcommand{\figref}[1]{Fig.~\ref{#1}}
\newcommand{\tabref}[1]{Tab.~\ref{#1}}
\newcommand{\secref}[1]{Sec.~\ref{#1}}
\newcommand{\appref}[1]{Appendix~\ref{#1}}

\begin{document}

\pagestyle{SPstyle}

\begin{center}{\Large \textbf{\color{scipostdeepblue}{
Unifying Dirac Spin Liquids on Square and\\ Shastry–Sutherland Lattices via Fermionic Deconfined Criticality
}}}\end{center}

\begin{center}\textbf{
Andreas~Feuerpfeil\,\orcidlink{0009-0001-0436-0332}\textsuperscript{1,2,$\star$},
Leyna~Shackleton\,\orcidlink{0000-0002-1971-6426}\textsuperscript{3},
Atanu~Maity\,\orcidlink{0000-0001-7822-124X}\textsuperscript{1},
Ronny~Thomale\,\orcidlink{0000-0002-3979-8836}\textsuperscript{1,4},
Subir~Sachdev\,\orcidlink{0000-0002-2432-7070}\textsuperscript{5,2} and 
Yasir~Iqbal\,\orcidlink{0000-0002-3387-0120}\textsuperscript{4,$\dagger$}
}\end{center}

\begin{center}
\textsuperscript{\bf 1}Institut für Theoretische Physik und Astrophysik and W\"urzburg-Dresden Cluster of Excellence ctd.qmat, Universit\"at W\"urzburg, 97074 W\"urzburg, Germany
\\
\textsuperscript{\bf 2}Center for Computational Quantum Physics, Flatiron Institute, 162 5th Avenue, New York, NY 10010, USA
\\
\textsuperscript{\bf 3}Department of Physics, Massachusetts Institute of Technology, Cambridge, Massachusetts
02139, USA
\\ 
\textsuperscript{\bf 4}Department of Physics,
Indian Institute of Technology Madras, Chennai 600036, India
\\
\textsuperscript{\bf 5}Department of Physics, Harvard University, Cambridge MA 02138, USA

$\star$ \href{mailto:andreas.feuerpfeil@uni-wuerzburg.de}{\small andreas.feuerpfeil@uni-wuerzburg.de}\,,\quad
$\dagger$ \href{mailto:yiqbal@physics.iitm.ac.in}{\small yiqbal@physics.iitm.ac.in}
\end{center}

\section*{\color{scipostdeepblue}{Abstract}}
\textbf{\boldmath{%
We present a fermionic gauge theory for deconfined quantum criticality on the Shastry–Sutherland lattice and reveal its shared low-energy field-theoretic structure with the square lattice. Starting from an $\SU(2)$ $\pi$-flux parent state, we construct a continuum theory of Dirac spinons coupled to an $\SU(2)$ gauge field and adjoint Higgs fields 
whose condensates drive transitions to a staggered-flux $\U(1)$ spin liquid and a gapless $\mathbb{Z}_{2}$ Dirac spin liquid. While the Shastry--Sutherland lattice permits additional symmetry-allowed fermion bilinears compared to the square lattice, the quantum field theories are identical up to additional irrelevant terms. Consequently, the Higgs potential structure and the leading low-energy theory coincide with the square-lattice case at the quantum critical point. The $\SO(5)$ critical point is expected to realize conformal deconfined criticality: we analyze it in a large-flavor expansion, calculate its critical exponents, and identify the Yukawa coupling between the fermions and Higgs fields as the relevant perturbation that destabilizes it, consistent with pseudocritical behavior observed in recent Monte Carlo studies. We show that the emergent $\SO(5)$ order parameter acquires a large anomalous dimension at the critical point, leading to strongly enhanced N\'eel and VBS susceptibilities---a hallmark of fermionic deconfined quantum criticality consistent with numerical studies. Our results place recent numerical evidence for a gapless $\mathbb{Z}_{2}$ Dirac spin liquid on the Shastry--Sutherland lattice within a controlled field-theoretic framework and demonstrate that fermionic deconfined criticality on the square lattice---including critical exponents and stability---extends to frustrated lattices with reduced symmetry.
}}

\vspace{\baselineskip}

\noindent\textcolor{white!90!black}{%
\fbox{\parbox{0.975\linewidth}{%
\textcolor{white!40!black}{\begin{tabular}{lr}%
  \begin{minipage}{0.6\textwidth}%
    {\small Copyright attribution to authors. \newline
    This work is a submission to SciPost Physics. \newline
    License information to appear upon publication. \newline
    Publication information to appear upon publication.}
  \end{minipage} & \begin{minipage}{0.4\textwidth}
    {\small Received Date \newline Accepted Date \newline Published Date}%
  \end{minipage}
\end{tabular}}
}}
}


\vspace{10pt}
\noindent\rule{\textwidth}{1pt}
\tableofcontents
\noindent\rule{\textwidth}{1pt}
\vspace{10pt}

\label{sec:introduction}
\noindent
Quantum phase transitions in two-dimensional antiferromagnets can fall outside the conventional Landau--Ginzburg--Wilson paradigm when the critical degrees of freedom are fractionalized and coupled to emergent gauge fields. A prominent example is deconfined criticality, where continuous transitions can occur between phases with distinct broken symmetries, such as N\'eel order and valence-bond-solid (VBS) order~\cite{Senthil2004b,Senthil2004a,Sandvik2007}. Considerable insight into such transitions has been obtained by formulating effective field theories of Dirac fermions coupled to gauge fields; this framework provides a unified description of magnetically ordered phases, spin liquids, and their proximate critical points~\cite{Hermele2005,Lee2006,SongPRX2020}.

Within this fermionic framework, $\SU(2)$ gauge theories of Dirac spinons play a central role~\cite{Affleck1988}. In these theories, adjoint Higgs fields reduce the gauge symmetry to $\U(1)$ or $\mathbb{Z}_2$, yielding staggered-flux~\cite{Lee2006} and gapless $\mathbb{Z}_2$ Dirac spin-liquid phases~\cite{Wen2002,Hermele2004,Hermele2007}, respectively. On the square lattice, such Higgs theories~\cite{Read-1991,Sachdev-1990} have been shown to capture deconfined critical points separating N\'eel order, VBS order, and gapless spin liquids, with the structure of the low-energy theory strongly constrained by lattice symmetries and their projective symmetry group (PSG) realizations~\cite{Tanaka2005,Senthil2006,Thomson_2018}.

The Shastry--Sutherland lattice provides a natural setting in which to examine these ideas beyond the square lattice. While retaining nontrivial point-group symmetries, it breaks several lattice symmetries of the square lattice and supports additional frustration through diagonal couplings~\cite{Shastry-1981}. Recently, numerical~\cite{Yang-2022,Wang-2022,Luciano-2024,Keles-2022,corboz2025,Chen2025} and variational studies~\cite{Maity2024} by some of us have provided evidence for a gapless $\mathbb{Z}_2$ Dirac spin liquid stabilized in the spin-$1/2$ Heisenberg model on this lattice, denoted Z3000; it can be interpreted as a descendant of a staggered-flux $\U(1)$ state~\cite{Lee2006} and, ultimately, of an $\SU(2)$ $\pi$-flux parent state originally proposed by Affleck and Marston for square-lattice antiferromagnets~\cite{Affleck1988,Affleck-SU2,Dagotto-1988,LeeWen96,Wen2002}. 

A central issue in extending fermionic deconfined criticality to the Shastry--Sutherland lattice is the role of reduced lattice symmetry. Whereas it might help to microscopically stabilize quantum paramagnets from the viewpoint of ground state energy density through enhanced tuning of frustration, it also enormously broadens the scope of symmetry-allowed ground state candidates: implementing the relevant PSG requires an enlarged $2 \times 2$ unit cell, which reorganizes the Dirac fermions and their symmetry transformations. While the square-lattice $\pi$-flux theory features Dirac valleys protected at distinct momenta, this protection is lifted in the Shastry--Sutherland embedding once lattice-specific perturbations are included. Furthermore, the reduced space-group symmetry enlarges the set of symmetry-allowed Yukawa couplings between fermion bilinears and Higgs fields, raising the question of whether the resulting continuum theory retains the same structure as in the square-lattice case. We show that while the reduced symmetry of the Shastry--Sutherland lattice allows for additional symmetry-allowed Yukawa couplings between Dirac fermions and Higgs fields, all such couplings share the same scaling dimension at leading order in a controlled large-$N$ expansion. Consequently, within this framework, lattice frustration does not modify the stability properties of the $\SO(5)$ critical point relative to the square-lattice theory.

The main results of this work are as follows. We construct a fermionic-spinon field theory for the Shastry--Sutherland model starting from the $\SU(2)$ $\pi$-flux parent state and derive the continuum perturbations that generate the $\U(1)$ staggered-flux and $\mathbb{Z}_2$ Dirac spin-liquid descendants. The resulting continuum description is a Dirac theory coupled to an $\SU(2)$ gauge field and three real adjoint Higgs fields, whose condensates encode the Higgs transitions between these phases. At the level of the most relevant couplings controlling the Higgs transitions, the resulting Majorana--Higgs theory coincides with the corresponding square-lattice theory, including a global $\SO(5)$ symmetry~\cite{Thomson_2018,Christos_2024}. The reduced symmetry of the Shastry--Sutherland lattice permits additional fermion bilinears and gradient terms, but these do not change the scaling dimensions or appear only as symmetry-allowed subleading perturbations. We further determine the most general Higgs potential consistent with the microscopic symmetries and show that its structure is identical to the square-lattice case to all orders, implying the same mean-field Higgs phase structure underlying the sequence N\'eel $\rightarrow$ $\mathbb{Z}_2$ Dirac spin liquid $\rightarrow$ VBS. Finally, we extract consequences for the competing N\'eel and VBS channels, including characteristic low-energy corrections to their susceptibilities.

In previous analyses for the square lattice~\cite{Shackleton2021,Shackleton:2022zzm}, it was assumed that the $\SO(5)$ breaking perturbations were strongly relevant, and this led to consideration of a critical theory without $\SO(5)$, or even Lorentz, symmetry. The present paper will take a complementary point of view, assuming that the $\SO(5)$ breaking perturbations are at best weakly relevant. We will focus on a theory with exact $\SO(5)$ symmetry, an $\SU(2)$ gauge field, $N_f=2$ massless, fundamental Dirac fermions and $N_b = 2$ adjoint Higgs fields. Studies of the N\'eel--VBS transition on the square lattice are described by $N_f=2$, $N_b=0$: state-of-the-art Monte Carlo studies increasingly indicate that this $\SO(5)$ point is pseudocritical or multifractal rather than truly conformal~\cite{Nahum2015,Wang_2017,WangPRL2021,Zhou_2024,Chen_2024,Meng24,Takahashi2024}, suggesting a wide pre-asymptotic regime with approximate $\SO(5)$ symmetry~\cite{Shao2016Science,Zhou_2024}. These results strongly imply the existence of a conformal fixed point with exact $\SO(5)$ symmetry for the $N_f=2$, $N_b=2$ case, as the additional critical Higgs fields can only enhance the stability of the CFT. We will study this CFT in the large-$N_{f,b}$ expansion, and compute scaling dimensions of various observables. 

Turning to the onset of the gapless $\mathbb{Z}_2$ Dirac spin liquid, we consider the perturbative influence of the Yukawa couplings which break both $\SO(5)$ and Lorentz symmetry present for this case (for both the square and Shastry--Sutherland lattices) at the fixed point with exact $\SO(5)$ symmetry.  
While our renormalization-group analysis of the associated gauge theory finds the leading $\SO(5)$-breaking perturbation to be relevant at large $N$, and hence to destabilize the $\SO(5)$-symmetric fixed point at the lowest energies, this does not contradict either earlier field-theoretic expectations or recent numerical work. First, the large-$N$ expansion does not control the fixed point at physical, finite $N$ in strongly coupled gauge theories, but rather correctly identifies the operator content~\cite{Sachdev2018}. Second, a defining and quantitatively testable signature of such fermionic deconfined critical points is the appearance of unusually large anomalous dimensions for the competing N\'eel and VBS order parameters. These have been observed numerically in $J$\textendash$Q$ models and related systems, and sharply distinguish deconfined criticality from conventional $\mathrm {O}(3)$ universality classes. In this work, we compute these anomalous dimensions analytically for the $\SU(2)$ $\pi$-flux theory relevant to both square and Shastry--Sutherland lattices. Taken together, these results suggest that while the gapless Dirac spinon theory on the Shastry--Sutherland lattice is ultimately unstable at the lowest energies, it can remain a valid and predictive description over an extended low-energy window, which may be relevant for experimentally accessible regimes.

The remainder of the paper is organized as follows. In Sec.~\ref{sec:meanfield} we introduce the fermionic spinon formulation and the Z3000 $\mathbb{Z}_2$ Dirac spin-liquid \textit{ansatz} on the Shastry--Sutherland lattice. Section~\ref{sec:continuum} derives the continuum $\SU(2)$ $\pi$-flux theory and the Higgs perturbations leading to the U800 and Z3000 states, and presents the resulting Majorana--Higgs Lagrangian and Higgs potential, together with a classification of all symmetry-allowed fermion bilinears up to first order in gradients under the Shastry--Sutherland space group. Section~\ref{sec:rg} presents a renormalization-group analysis of the resulting gauge theory including a calculation of the critical exponents, and Sec.~\ref{sec:conclusion} discusses the implications for deconfined criticality in frustrated quantum magnets.

\section{Gapless \texorpdfstring{$\bm{\mathbb{Z}_2}$}{Z2} spin liquid on the Shastry--Sutherland lattice}\label{sec:meanfield}

\begin{figure}
    \centering
    \includegraphics[width=0.75\linewidth]{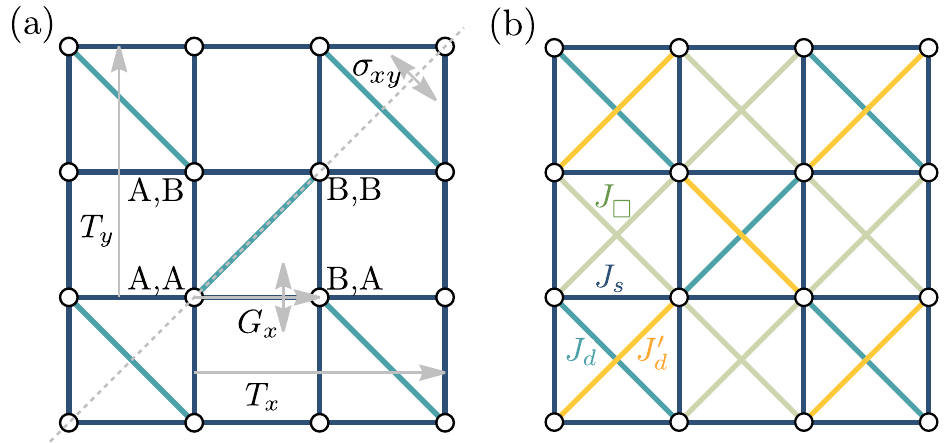}
	\caption{(a) Illustration of the Shastry--Sutherland lattice. The four orbitals within the unit cell are denoted by two sublattice degrees of freedom $(m_x,m_y)=(A,A), (B,A), (B,B), (A,B)$, respectively. (b) Shown are additional diagonal links with the Heisenberg couplings needed to interpolate to the square lattice limit.}
	\label{fig:lattice}
\end{figure}

The Shastry--Sutherland Hamiltonian for a spin-$1/2$ Heisenberg antiferromagnet is defined on the lattice, see \figref{fig:lattice}, as
\begin{equation}
\label{eq:model}
    \hat{H} = J_s\sum_{\langle \vi,\vj\rangle_{\mathrm{square}}}\hatbold{S}_\vi\cdot\hatbold{S}_\vj +J_d\sum_{\langle \vi,\vj\rangle_\mathrm{dimer}}\hatbold{S}_\vi\cdot\hatbold{S}_\vj\,.
\end{equation}
where $\hat{\mathbf{S}}_\vi$ denotes the spin operator at site $\vi$, and $J_{s}$ and $J_d$ are the exchange couplings on the edges of the squares and on the dimer bonds, respectively (see Fig.~\ref{fig:lattice}(b)). Equation~\eqref{eq:model} accurately models the spin-$\tfrac12$ Shastry--Sutherland compound SrCu$_2$(BO$_3$)$_2$, which realizes an exact dimer-singlet ground state at ambient pressure~\cite{Kageyama1999PRL}. Hydrostatic pressure tunes the effective ratio $J_d/J_s$, driving a sequence of quantum phase transitions resolved by thermodynamic, spectroscopic, and neutron-scattering experiments, including an intermediate gapped plaquette phase followed by antiferromagnetic order~\cite{Haravifard2014PNAS,Zayed2017NatPhys}. In the fermionic gauge-theory framework, increasing $J_d/J_s$ corresponds to Higgs-induced reductions of the emergent $\SU(2)$ gauge structure, whose confined descendants naturally account for plaquette and N\'eel phases, providing a materials setting for pressure-driven deconfined or pseudocritical behavior~\cite{Cui-2023,Guo-2025}.

We therefore formulate the fermionic spinon theory for $\mathbb{Z}_2$ spin liquids~\cite{Wen-1991} on the Shastry--Sutherland lattice. We begin by reexpressing the spin operators with respect to spinons $f_{\vi\alpha}, \alpha=\uparrow, \downarrow$ for sites $\vi=(i_x,i_y)$~\cite{Abrikosov-1965}:
\begin{equation}
    \mbf{S}_{\vi}=\frac{1}{2}f_{\vi\alpha}^\dag \bm{\sigma}_{\alpha\beta}f_{\vi\beta}\,.
\end{equation}
Following Wen~\cite{Wen2002}, we introduce a Nambu spinor 
\begin{equation}
    \psi_{\vi}=\begin{pmatrix}
        f_{\vi\uparrow}\\f_{\vi\downarrow}^\dag
    \end{pmatrix}\,,
\end{equation}
which leads to the Bogoliubov Hamiltonian
\begin{equation}\label{eq:bogoliubov_hamiltonian}
    H=-\sum_{\vi\vj}\psi_{\vi}^\dag u_{\vi\vj}\psi_{\vj}\,.
\end{equation}
Here, 
\begin{equation}
    u_{\vi \vj} = iu_{\vi\vj}^0\tau^0+u_{\vi\vj}^a\tau^a\,,
\end{equation}
with the Pauli matrices $\tau^\mu$ acting on the Nambu spinor $\psi_{\vi}$. Spin-rotation symmetry implies that $u_{\vi\vj}^\mu \in \mathbb{R}$ and 
\begin{equation}
    u_{\vj\vi}^0=-u_{\vi\vj}^0, \quad  u_{\vj\vi}^a=u_{\vi\vj}^a\,.
\end{equation}
The spinon representation is invariant under $\SU(2)_g$ gauge transformations~\cite{Affleck1988,Dagotto-1988} for a symmetry operator $g$:
\begin{equation}
\begin{split}
    \SU(2)_g: \psi_{\vi}&\rightarrow U_{g,g(\vi)}\psi_{g(\vi)}\,,\\
    u_{\vi\vj}&\rightarrow U_{g,g(\vi)}u_{g(\vi),g(\vj)}U^\dag_{g,g(\vj)}\,.
\end{split}    
\end{equation}

Based on a recent variational study~\cite{Maity2024} by some of us and others, a relevant candidate for the spin-liquid phase of the Shastry--Sutherland model is the stable $\mathbb{Z}_2$ Dirac spin liquid Z3000, which can be understood as a descendant of the U800 staggered flux state, which derives from the $\SU(2)$ $\pi$-flux state of the square lattice~\cite{Wen2002}. As the $\mathbb{Z}_2$ \textit{ansatz} breaks the square lattice symmetries down to the Shastry--Sutherland lattice, we introduce two sublattice degrees of freedom $m_x,m_y=A,B$ labeling the positions in the $2\times 2$ unit cell as shown in \figref{fig:lattice}. We denote the Pauli matrices acting on $m_x$ as $\rho^\mu$, and the ones acting on $m_y$ as $\kappa^\mu$. In the following, when writing $\vi$ or $\vj$, we reference the lattice site, whilst when writing $\vecr$, we reference a $2\times2$ unit cell.

Explicitly, the \textit{ansatz} labeled by Z3000 has the following structure:
\begin{align}
u^{}_{\mbf{r},A,A,\mbf{r},B,A}&=
\begin{bmatrix}
-t e^{-\dot\iota\theta} &0\\
0 & t e^{\dot\iota\theta}
\end{bmatrix}=u^{}_s\,,\label{eq:U1_1}\\
u^{}_{\mbf{r},A,A,\mbf{r},A,B}&=
\begin{bmatrix}
t e^{\dot\iota\theta} &0\\
0 & -t e^{-\dot\iota\theta}
\end{bmatrix}=-u^{\dagger}_s\,,\\
u^{}_{\mbf{r},A,B,\mbf{r},B,B}&=u^{}_{\mbf{r},B,B,\mbf{r},B,A}=u^{}_{\mbf{r},B,A,\mbf{r}+\mbf{\hat{x}},A,A}=-u^{\dagger}_s\,,\\
u^{}_{\mbf{r},A,B,\mbf{r}+\mbf{\hat{y}},A,A}&=u^{}_{\mbf{r}+\mbf{\hat{y}},B,A,\mbf{r},B,B}=u^{}_{\mbf{r},B,B,\mbf{r}+\mbf{\hat{x}},A,B}=u^{}_s\label{eq:U1_4}\,,\\
u^{}_{\mbf{r},A,A,\mbf{r},B,B}&=\Delta^{}_d\tau^x,\;
u^{}_{\mbf{r}+\mbf{\hat{x}},A,B,\mbf{r}+\mbf{\hat{y}},B,A}=\Delta^{}_d \tau^y \label{eq:Z2_1}\,,\\
u^{}_{\mbf{r},B,B,\mbf{r}+\mbf{\hat{x}}+\mbf{\hat{y}},A,A}&=\Delta^{}_{d'}\tau^x,\;
u^{}_{\mbf{r},B,A,\mbf{r},A,B}=\Delta^{}_{d'}\tau^y\,,\\
u^{}_{\mbf{r},B,A,\mbf{r}+\mbf{\hat{x}},A,B}&=-\Delta_{g,2}\tau^x+\Delta_{g,1}\tau^y\,,\\
u^{}_{\mbf{r},A,B,\mbf{r}+\mbf{\hat{y}},B,A}&=\Delta_{g,2}\tau^x+\Delta_{g,1}\tau^y\,,\\
u^{}_{\mbf{r}+\mbf{\hat{x}},A,A,\mbf{r},B,B}&=\Delta_{g,1}\tau^x+\Delta_{g,2}\tau^y\,,\\
u^{}_{\mbf{r},B,B,\mbf{r}+\mbf{\hat{y}},A,A}&=\Delta_{g,1}\tau^x-\Delta_{g,2}\tau^y\,. \label{eq:Z2_end}
\end{align}
Equations~\eqref{eq:U1_1}--\eqref{eq:U1_4} describe the fermionic hopping terms, while the remaining equations correspond to pairing terms. Depending on the parameters, this \textit{ansatz} yields a $\mathbb{Z}_2$ spin liquid or its parent $\U(1)$ and $\SU(2)$ states:
\begin{enumerate}[label=\roman*.]
\item Setting the pairing terms to zero ($\Delta=0$) and fixing $\theta=\pi/2$ yields the parent $\pi$-flux state with $\SU(2)$ gauge structure. This state is labeled SU2B$n$0 in Ref.~\cite{Wen2002}.
\item Vanishing pairing ($\Delta=0$) with generic $\theta$ leads to a $\U(1)$ state with a staggered $(\varphi,-\varphi)$ flux through the square plaquettes, where $\varphi=\pi+4\theta$. This state corresponds to U800 in Ref.~\cite{Maity2024}.
\item Introducing finite pairing terms ($\Delta \neq 0$) breaks the $\U(1)$ gauge symmetry of the U800 state down to $\mathbb{Z}_2$, resulting in the $\mathbb{Z}_2$ Dirac spin liquid labeled Z3000~\cite{Maity2024}.
\end{enumerate}
These hopping and pairing mean-field parameters are illustrated in \figref{fig:u800_z3000_ansatz}, where the imaginary and real components of the hopping are defined as $t_{s,0}=t \sin\theta$ and $t_{s,z}=t \cos\theta$, respectively. We discuss the stability of the Z3000 spin liquid in \appref{app:PSGs_Z3000}.

\section{Continuum theory for Higgs transition from \texorpdfstring{$\bm{\SU(2)}$}{SU(2)} to \texorpdfstring{$\bm{\mathbb{Z}_2}$}{Z2}}
\label{sec:continuum}
\subsection[\texorpdfstring{${\SU(2)}$ ${\pi}$}{SU(2) pi}-flux state]{\texorpdfstring{$\bm{\SU(2)}$ $\bm{\pi}$}{SU(2) pi}-flux state} \label{sec:su2_continuum}
We begin by working out the $\SU(2)$ gauge theory based on the PSG \textit{ansatz} for the $\pi$-flux state~\cite{Affleck1988,Wen2002}. We replace the Nambu spinor by a matrix operator of Majorana fermions

\begin{equation}\label{eq:majorana_matrix_operator}
    \mathcal{F}_{\vi}=\begin{pmatrix}
        f_{\vi\uparrow} && -f_{\vi\downarrow}^\dag \\
        f_{\vi\downarrow} && f_{\vi\uparrow}^\dag
    \end{pmatrix}\,.
\end{equation}

This matrix operator satisfies the reality condition
\begin{equation}\label{eq:reality_condition}
    \mathcal{F}_{\vi}^\dag=\tau^y\mathcal{F}_{\vi}^T \tau^y \,.
\end{equation}
$\SU(2)_g$ gauge transformations $U_g$ correspond to left multiplication 
\begin{equation}
    \mathcal{F}_{\vi}\rightarrow U_{g,g(\vi)}\mathcal{F}_{g(\vi)}\,,
\end{equation}
whilst $\SU(2)$ spin rotations $\Omega$ correspond to the right multiplication
\begin{equation}
    \mathcal{F}_{\vi}\rightarrow \mathcal{F}_{\vi}\tau^z\Omega_{\vi}^T\tau^z\,.
\end{equation}
The Hamiltonian \eqnref{eq:bogoliubov_hamiltonian} then reads
\begin{equation}\label{eq:H_MF_matrix_operator}
\begin{split}
H=\sum_{\langle\vi\vj\rangle}&i\alpha_{\vi\vj}\Tr[\mathcal{F}_{\vi}^\dag \mathcal{F}_{\vj}]+\beta_{\vi\vj}^a\Tr[\tau^a\mathcal{F}_{\vi}^\dag \mathcal{F}_{\vj}] + i\gamma_{\vi\vj}\Tr[\tau^a\mathcal{F}_{\vi}^\dag \tau^a\mathcal{F}_{\vj}]
\end{split}
\end{equation}
with 
\begin{equation}
    u_{\vi\vj}=i\alpha_{\vi\vj}\tau^0+\beta_{\vi\vj}^a \tau^a\, .
\end{equation}
The additional $\gamma_{\vi\vj}$ hoppings involve projective realizations of the spin rotation symmetry~\cite{Chen-2012} and will not be relevant in the following.

The $\pi$-flux state has hoppings $\beta^a=0$ and
\begin{equation}
\begin{split}
    \alpha_{\vi\vj}&=-\alpha_{\vj\vi}, \quad \alpha_{\vi,\vi+\hatx}=t, \quad \alpha_{\vi,\vi+\haty}=(-1)^{i_x}t \,.
\end{split}
\end{equation}
This \textit{ansatz} respects the square lattice symmetries and is, up to a gauge transformation, identical to the \textit{ansatz} studied in Ref.~\cite{Shackleton2021}. To study the low-energy physics, we express the Hamiltonian in momentum space
\begin{equation}
    H=-2t\sum_{k}\Tr[\mathcal{F}^\dag_\vk\left(\sin\left(k_x/2\right)\rho^x+\sin\left(k_y/2\right)\rho^z\kappa^x\right)\mathcal{F}_\vk]\,,
\end{equation}
where the summation over $m_x,m_y$ is implicit. The Hamiltonian features a Dirac point at $\Gamma=(0,0)$ with a four-fold degeneracy arising from the sublattice degrees of freedom. The dispersion of the U800 spin liquid is shown in \figref{fig:u800_dispersion}, where $\varphi=\pi$ corresponds to the $\SU(2)$ $\pi$-flux state.

In comparison, on the square lattice, introducing the sublattice degree of freedom $m_x$ is sufficient to restore translational invariance of the \textit{ansatz}. This yields two Dirac points associated with a valley degree of freedom $v=0,1$ at $k_x=0$ and $k_y = 0,\pi$. These points are back-folded to the $\Gamma$ point due to the doubling of the unit cell in the $y$-direction. To make the valley degree of freedom explicit, we introduce two low-energy, two-component fermions $\X_{\vecr,m_x,v}$:
\begin{equation}\label{eq:low_energy_modes}
    \mathcal{F}_{\vecr,m_x,m_y}=\sum_{m_x'}\rho^x_{m_x,m_x'}\X_{\vecr,m_x',0}+\kappa^z_{m_y,m_y}\X_{\vecr,m_x,1}\,.
\end{equation}
The fields $\X_{\vecr,m_x,v}$ introduced in \eqnref{eq:low_energy_modes} are not independent complex Dirac fermions, but low-energy components of the Majorana matrix field $\mathcal{F}_{\vecr,m_x,m_y}$ defined in \eqnref{eq:majorana_matrix_operator}. Since \eqnref{eq:low_energy_modes} only rewrites $\mathcal{F}$ as a linear combination of low-energy modes in the sublattice/valley sector, with real coefficients given by $\rho^x$ and $\kappa^z$, the Majorana reality condition
\begin{equation}
\mathcal{F}^\dagger = \tau^y \mathcal{F}^T \tau^y
\end{equation}
is inherited by the low-energy fields $\X_{\vecr,m_x,v}$. Equivalently, each $\X_{\vecr,m_x,v}$ can be written in the same Nambu form as $\mathcal{F}$, namely
\begin{equation}
\X_{\vecr,m_x,v}
=
\begin{pmatrix}
\chi_{\vecr,m_x,v,\uparrow} & -\chi^\dagger_{\vecr,m_x,v,\downarrow}\\
\chi_{\vecr,m_x,v,\downarrow} & \chi^\dagger_{\vecr,m_x,v,\uparrow}
\end{pmatrix}\,,
\qquad
\X_{\vecr,m_x,v}^\dagger = \tau^y \X_{\vecr,m_x,v}^T \tau^y,
\end{equation}
where $\chi_{\vecr,m_x,v,\alpha}$ denote the corresponding low-energy spinon operators. Thus, \eqnref{eq:low_energy_modes} does not introduce a new Nambu structure, but only expresses the original Majorana field in the low-energy valley basis.

We introduce the Pauli matrices $\mu^a$ acting on the valley degree of freedom. In the square lattice model---and by extension, the $\mathrm{SU}(2)$ $\pi$-flux state of the Shastry--Sutherland model---the valleys reside at distinct momenta, which forbids intervalley scattering. However, this protection is eventually lifted by perturbations due to backfolding effects in the Shastry--Sutherland model. After explicitly verifying that there are no cross terms between $v=0$ and $1$, we find the Hamiltonian to leading order:
\begin{equation}
\begin{split}
    H\approx it\sum_{v}\Tr[\X_{v}^\dag (\rho^x\partial_x-\rho^z\partial_y) \X_{v}]\,.
\end{split}
\end{equation}
Finally, we obtain a relativistic Dirac equation as the continuum Lagrangian
\begin{equation}
\begin{split}
    \mathcal{L}_{\mathrm{MF}}&=i\Tr[\bar{\X}\gamma^\mu \partial_\mu\X]\,,
\end{split}
\end{equation}
where we absorbed $t$ into the speed of light and used the Lorentz signature $(+,-,-)$ with $\bar{\X}=\X^\dag\gamma^0$ and the gamma matrices $\gamma^0=\rho^y,\gamma^x=i\rho^z,\gamma^y=i\rho^x$. 

This theory has an emergent global $\SU(4)$ symmetry acting on the four-dimensional space of spin ($\sigma$) and valley ($\mu$)~\cite{Thomson_2018,Shackleton2021}. Imposing the reality condition \eqnref{eq:reality_condition} reduces this $15$-dimensional $\SU(4)$ symmetry down to its $10$-dimensional subgroup $\Sp(4)$. Since $\Sp(4)$ and $\SU(2)_g$ share the element $-1$, the global symmetry is given by $\Sp(4)/\mathbb{Z}_2\cong \SO(5)$. The group action is defined by
\begin{equation}
    \X\rightarrow \X L\,,
\end{equation}
for which the reality condition implies
\begin{equation}
    L^T=\tau^y L^\dag \tau^y\,.
\end{equation}
For the Lie generators $M=M^\dag,L=e^{iM}$, this is equivalent to 
\begin{equation}\label{eq:reality_condition_lie}
    M^T=-\tau^yM\tau^y\,.
\end{equation}
The $10$ generators of $\SU(4)$ that satisfy the reality condition are given by
\begin{equation}
    T^j=\lbrace \mu^y, \tau^a,\mu^x\tau^a,\mu^z\tau^a\rbrace\,,
\end{equation}
whilst the remaining $5$ $\SU(4)$ generators that do not satisfy \eqnref{eq:reality_condition_lie} all anti-commute and are given by 
\begin{equation}
    \Gamma^j =\lbrace\mu^x,\mu^z,\mu^y\tau^a\rbrace\,.
\end{equation}
They transform as a vector under the $\SO(5)$ symmetry. The corresponding fermion bilinear $i\Tr[\bar{\X}\Gamma^j \X]$ combines the VBS and N\'eel order parameter~\cite{Tanaka2005,Ran2006,Wang_2017}.

In the following subsections, we derive the continuum Lagrangian for the perturbations in Eqs.~\eqref{eq:U1_1}--\eqref{eq:Z2_end}, which reduce the $\pi$-flux state to the U800 staggered-flux or Z3000 spin-liquid phases. We perform a gradient expansion of these perturbations using the low-energy modes defined in \eqnref{eq:low_energy_modes}. By coupling these terms to adjoint Higgs fields, we capture the phase transitions to the $\U(1)$ or $\mathbb{Z}_2$ spin liquids via Higgs condensation~\cite{Read-1991}.
\subsection[\texorpdfstring{From ${\pi}$-flux to U800}{From pi-flux to U800}]{\texorpdfstring{From $\bm{\pi}$-flux to U800}{From pi-flux to U800}}\label{sec:gaplessU1}
We obtain the continuum version of the perturbations to the U800 spin liquid by expanding around the mean field $u_{\vi\vj}$ in powers of $\theta=\pi/2+\delta \theta$ corresponding to additional hopping parameters
\begin{equation}
    \beta_{\vi,\vi+\hatx}^z=-\delta\theta(-1)^{i_x+i_y}\,, \quad \beta_{\vi,\vi+\haty}^z=\delta\theta(-1)^{i_y}\,.
\end{equation}
We start with the hopping in the $x$-direction, where we leave the sublattice indices implicit and reintroduce the trace and $\tau^z$ in the Lagrangian:
\begin{equation}
\begin{split}\label{eq:U1_perturbation_x}
    \delta H = &-\delta\theta\sum_{\vecr} [\X^\dag_{\vecr,0}\rho^x+\X^\dag_{\vecr,1}\kappa^z] \rho^x\kappa^z [\rho^x\X_{\vecr,0}+\kappa^z\X_{\vecr,1}]\\
    &+\frac{1}{2}\delta\theta \sum_{\vecr}[\X^\dag_{\vecr,0}\rho^x+\X^\dag_{\vecr,1}\kappa^z] (\rho^x-i\rho^y)\kappa^z [\rho^x\X_{\vecr+\hatx,0}+\kappa^z\X_{\vecr+\hatx,1}]\\
    &+\frac{1}{2}\delta\theta \sum_{\vecr}[\X^\dag_{\vecr,0}\rho^x+\X^\dag_{\vecr,1}\kappa^z] (\rho^x+i\rho^y)\kappa^z [\rho^x\X_{\vecr-\hatx,0}+\kappa^z\X_{\vecr-\hatx,1}] \\
    \approx &\, 2\delta\theta \int \dd^2 \vecr [\X^\dag_{\vecr,0}\rho^z\partial_x\X_{\vecr, 1}-\X_{\vecr,1}^\dag\rho^z\partial_x\X_{\vecr,0}]\\
    \Rightarrow \delta \mathcal{L} =&-2i\delta\theta \Tr[\tau^z\bar{\X}\mu^y\gamma^y\partial_x \X]\,.
\end{split}
\end{equation}
For the hopping in the $y$-direction,
\begin{equation}
\begin{split}\label{eq:U1_perturbation_y}
    \delta H = &+\delta\theta\sum_{\vecr} [\X^\dag_{\vecr,0}\rho^x+\X^\dag_{\vecr,1}\kappa^z] \kappa^x [\rho^x\X_{\vecr,0}+\kappa^z\X_{\vecr,1}]\\
    &-\frac{1}{2}\delta\theta \sum_{\vecr}[\X^\dag_{\vecr,0}\rho^x+\X^\dag_{\vecr,1}\kappa^z] (\kappa^x-i\kappa^y) [\rho^x\X_{\vecr+\haty,0}+\kappa^z\X_{\vecr+\haty,1}]\\
    &-\frac{1}{2}\delta\theta \sum_{\vecr}[\X^\dag_{\vecr,0}\rho^x+\X^\dag_{\vecr,1}\kappa^z] (\kappa^x+i\kappa^y) [\rho^x\X_{\vecr-\haty,0}+\kappa^z\X_{\vecr-\haty,1}]\\
    \approx &\, -2\delta\theta \int \dd^2 \vecr [\X^\dag_{\vecr,0}\rho^x\partial_y\X_{\vecr,1}-\X_{\vecr,1}^\dag\rho^x\partial_y\X_{\vecr,0}]\\
    \Rightarrow \delta \mathcal{L} =&-2i\delta\theta \Tr[\tau^z\bar{\X}\mu^y\gamma^x\partial_y \X]\,.
\end{split}
\end{equation}
In both \eqnref{eq:U1_perturbation_x} and \eqnref{eq:U1_perturbation_y} the Pauli matrix $\tau^z$ is acted on by the $\SU(2)$ gauge symmetry of the $\pi$-flux phase. Gauge invariance hence requires the existence of similar terms with $\tau^x$ and $\tau^y$. Therefore, we express the perturbation in a gauge-independent fashion by introducing an adjoint Higgs field $\Phi_3^a$ with $\SU(2)$ gauge index $a$:
\begin{equation}
    \delta \mathcal{L}=\Phi_3^a\Tr[\tau^a \bar{\X}\mu^y(\gamma^y i\partial_x+\gamma^x i\partial_y) \X]\,.
\end{equation}
The subscript ``3'' is chosen in accordance with~\cite{Shackleton2021}. The continuum version of \eqnref{eq:U1_perturbation_x} and \eqnref{eq:U1_perturbation_y} is obtained by condensing $\langle \Phi_3^z \rangle \propto \delta\theta$.

\subsection[{\texorpdfstring{From ${\pi}$-flux to Z3000}{From pi-flux to Z3000}}]{\texorpdfstring{From $\bm{\pi}$-flux to Z3000}{From pi-flux to Z3000}}\label{sec:gaplessZ2}
We now turn to the pairing terms $\Delta_d,\Delta_{d'},\Delta_{g,1/2}$ in Eqs.~\eqref{eq:Z2_1}--\eqref{eq:Z2_end}, which break the invariant gauge group down to a $\mathbb{Z}_2$ gauge group corresponding to the gapless Z3000 spin liquid~\cite{Maity2024}. The perturbation due to $\Delta_d$ is given by:
\begin{align}
    \delta H&=\Delta_d \Tr\Big\lbrace\sum_{\vecr} \tau^x[\X_{\vecr,B,0}^\dag+\X_{\vecr,A,1}^\dag][\X_{\vecr,A,0}-\X_{\vecr,B,1}]+\mathrm{h.c.}\nonumber\\
    &\qquad \qquad +\tau^y[\X_{\vecr,B,0}^\dag-\X_{\vecr,A,1}^\dag][\X_{\vecr-\hatx+\haty,A,0}+\X_{\vecr-\hatx+\haty,B,1}]+\mathrm{h.c.}\Big\rbrace\nonumber\\
    &\approx\Delta_d\int \dd^2\vecr \Tr\lbrace\tau^x(\X^\dag \rho^x\mu^z\X+\X^\dag \rho^z\mu^x\X)\rbrace+\Tr\lbrace\tau^y(\X_\vecr^\dag \rho^x\mu^z\X_\vecr-\X_\vecr^\dag \rho^z\mu^x\X_{\vecr})\rbrace\nonumber\\
    \Rightarrow \delta \mathcal{L}&=\Delta_d\Tr[\tau^x\bar{\X}\left(\gamma^x\mu^z-\gamma^y\mu^x\right)\X]+\Delta_d\Tr[\tau^y\bar{\X}\left(\gamma^x\mu^z+\gamma^y\mu^x\right)\X]
\end{align}
$\Delta_{d'}$ leads to the same term in the Lagrangian to first order and only differs from $\Delta_d$ by a less relevant gradient term, which we will ignore in the following.

Similarly, we can deal with $\Delta_{g,1/2}$. As $\Delta_{g,2}$ vanishes to lowest order, we also calculate the gradient term in $\Delta_{g,2}$, see \appref{app:pairing_terms}. In summary, we obtain the following Lagrangian:
\begin{equation}
\begin{split}\label{eq:Z2_perturbations}
    \delta \mathcal{L}&=(\Delta_d+\Delta_{d'}+2\Delta_{g,1})\left(\Tr[\tau^x\bar{\X}\left(\gamma^x\mu^z-\gamma^y\mu^x\right)\X]+\Tr[\tau^y\bar{\X}\left(\gamma^x\mu^z+\gamma^y\mu^x\right)\X]\right)\\
    &+\Delta_{g,2}\left(\Tr[\tau^x\bar{\X}(-\gamma^0\mu^y+1)(i\partial_x+i\partial_y)\X]+\Tr[\tau^y\bar{\X}(\gamma^0\mu^y+1)(i\partial_y-i\partial_x)\X]\right)
\end{split}
\end{equation}

Again, we want to express the Lagrangian in a gauge-invariant fashion. The first two terms are identical to the square lattice case~\cite{Shackleton2021}, for which one can introduce two Higgs fields $\Phi_{1,2}^a$ with the couplings
\begin{equation}
    \Phi_1^a\Tr[\tau^a \bar{\X}\gamma^x\mu^z\X]+\Phi_2^a\Tr[\tau^a \bar{\X}\gamma^y\mu^x\X]\,.
\end{equation}
In addition to these, we obtain two gradient terms due to the $\Delta_{g,2}$ pairing, which we can also absorb into these two Higgs fields due to them having the same symmetry transformations (see \tabref{tab:SS_bilinears_1} to \tabref{tab:SS_bilinears_3}) leading to the terms
\begin{equation}
    \Phi_1^a\Tr[\tau^a \bar{\X}(\gamma^x\mu^z-\delta(\gamma^0\mu^yi\partial_x-i\partial_y))\X]
\end{equation}
and 
\begin{equation}
    \Phi_2^a\Tr[\tau^a\bar{\X}(\gamma^y\mu^x+\delta(\gamma^0\mu^yi\partial_y-i\partial_x))\X]\,,
\end{equation}
where we defined the dimensionless scale $\delta \equiv \Delta_{g,2}/(\Delta_d+\Delta_{d'}+2\Delta_{g,1})$. We note in passing that the other pairings also lead to gradient terms, which we did not compute in this work, as they will not be relevant for the renormalization group study. Their effect would add coupling terms similar to $\delta$. The continuum version of \eqnref{eq:Z2_perturbations} is then obtained by condensing the Higgs fields as
\begin{equation}
\begin{split}
    \langle \Phi_1\rangle &\propto (\Delta_{d}+\Delta_{d'}+2\Delta_{g,1})(1,1,0)\,,\\
    \langle \Phi_2\rangle &\propto (\Delta_{d}+\Delta_{d'}+2\Delta_{g,1})(-1,1,0)\,.\\
\end{split}
\end{equation}

\subsection{Majorana-Higgs Lagrangian}
\label{sec:maj_higgs}

\begin{figure}[h]
    \centering
    \includegraphics[width=0.5\linewidth]{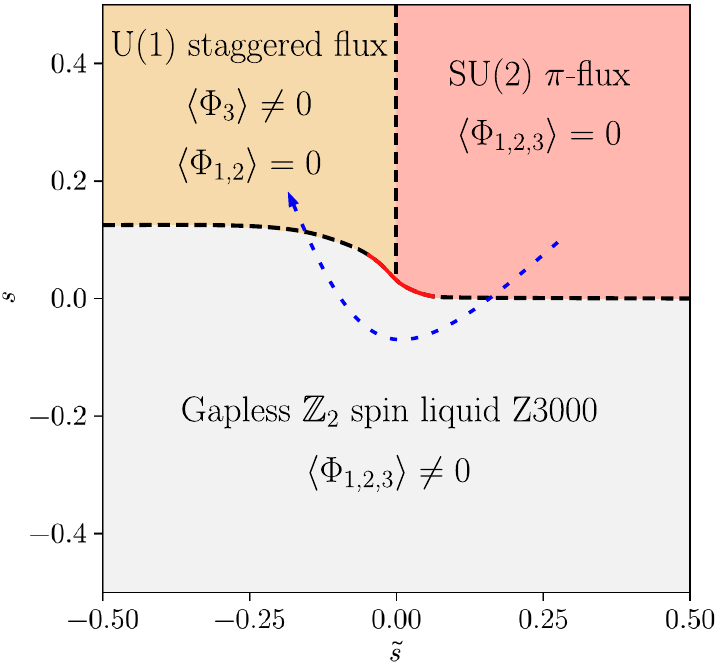}
    \caption{Mean-field phase diagram of the Higgs potential in \eqnref{eq:Higgs_potential}. The dashed (solid red) lines indicate second-order (first-order) transitions. Following~\cite{Shackleton2021}, we choose the mean-field parameters to be $w=u=1$, $v_2=-1$, $\tilde{u}=0.75$, and $v_4=0.5$. We employ the ansatz $\Phi_1^a = c_1 \delta_{a,x}$, $\Phi_2^a = c_1 \delta_{a,y}$, and $\Phi_3^a = c_2 \delta_{a,z}$. We assume that the $\SU(2)$ $\pi$-flux state confines to a N\'eel state whilst the $\U(1)$ staggered flux state confines to a columnar or plaquette VBS state. In case the confinements are opposite to our assumption, the direction of the arrow, which indicates a possible trajectory of the Shastry--Sutherland model with increasing $J_d/J_s$, has to be reversed. The $\mathbb{Z}_2$ spin liquid Z3000 is a stable Dirac spin liquid~\cite{Wen2002,Senthil-2000}. Our field-theoretic analysis shows that it is connected to Wen's~\cite{Wen2002} gapless spin liquid Z2A$zz$13, consistent with an infinite density matrix renormalization group, exact diagonalization and PSG/variational Monte Carlo study~\cite{Maity2024}.}
    \label{fig:mean_field_diagram}
\end{figure}

Now, we collect the perturbations to obtain the Lagrangian for the Majorana field $\mathcal{X}$ and three real adjoint Higgs fields $\Phi_{1,2,3}^a$. We will not explicate the $\SU(2)$ gauge fluctuations, as they can be added via minimal coupling:
\begin{equation}
\begin{split} \label{eq:full_lagrangian}
    \mathcal{L}=&i\Tr[\bar{\X}\gamma^\mu \partial_\mu \X]+\Phi_1^a\Tr[\tau^a\bar{\X}(\gamma^x\mu^z-\delta(\gamma^0\mu^yi\partial_x-i\partial_y)) \X]\\
    &+\Phi_2^a\Tr[\tau^a\bar{\X}(\gamma^y\mu^x+\delta(\gamma^0\mu^yi\partial_y-i\partial_x)) \X]\\
    &+\Phi_3^a\Tr[\tau^a \bar{\X}\mu^y(\gamma^yi\partial_x+\gamma^xi\partial_y)\X]+V(\Phi)\,.\\
\end{split}
\end{equation}
We deduce the form of the Higgs potential $V(\Phi)$, which arises from integrating out the high energy spinon degrees of freedom, by constructing all gauge-invariant and symmetry-allowed terms on the Shastry--Sutherland lattice up to quartic order from \tabref{tab:higgs_transformations}:
\begin{table}[htb]
    \centering
    \begin{tblr}{
          width = \linewidth,
          colspec = {c X[c] X[c] X[c] X[c] X[c] | X[c] X[c] X[c]},
          cell{2}{7,8,9} = {bg=colone},
          cell{3}{7,8,9}  = {bg=coltwo},
          cell{4}{7,8,9}  = {bg=colthree},
        }\hline\hline
         & $T_x$ & $T_y$ & $P_x$ & $P_y$ & $R_{\pi/2}$ & $G_x$ & $\mathcal{T}$ & $ \sigma_{xy}$\\\hline
         $\Phi_1^a$ & $-$ & $+$ & $-$ & $-$ & $-\Phi_2^a$ & $+$ &$-$ & $\Phi_2^a$ \\
         $\Phi_2^a$ & $+$ & $-$ & $-$ & $-$ & $-\Phi_1^a$ & $-$ & $-$ & $\Phi_1^a$\\
         $\Phi_3^a$ & $-$ & $-$ & $+$ & $+$ & $-$ & $-$ & $+$ & $-$\\\hline
    \end{tblr}
    \caption{Symmetry transformations of the Higgs fields under the square and Shastry--Sutherland lattice symmetries. Time reversal symmetry is implemented identically on the square and Shastry--Sutherland lattice. The transformation of the Higgs fields under it is identical on both lattices, so we show it only once, in the second-to-last column.}
    \label{tab:higgs_transformations}
\end{table}
\begin{equation}
\begin{split}\label{eq:Higgs_potential}
    V(\Phi)&=s(\Phi_1^a\Phi_1^a+\Phi_2^a\Phi_2^a)+\tilde{s}\Phi_3^a\Phi_3^a+w\epsilon_{abc}\Phi_1^a\Phi_2^b\Phi_3^c\\
    &+u(\Phi_1^a\Phi_1^a+\Phi_2^a\Phi_2^a)^2+\tilde{u}(\Phi_3^a\Phi_3^a)^2 +v_1 (\Phi_1^a\Phi_2^a)^2\\
    &+v_2 (\Phi_1^a\Phi_1^a)(\Phi_2^b\Phi_2^b)+ v_3[(\Phi_1^a\Phi_3^a)^2+(\Phi_2^a\Phi_3^a)^2] \\
    &+v_4(\Phi_1^a\Phi_1^a+\Phi_2^a\Phi_2^a)(\Phi_3^b\Phi_3^b)\,,
\end{split}
\end{equation}
where $\epsilon_{abc}$ is the Levi-Civita symbol. These terms are again identical to the square lattice case~\cite{Shackleton2021} and lead to a mean-field phase diagram with three different phases as exemplified in \figref{fig:mean_field_diagram}.

In the following, we prove that the symmetry-allowed terms in the Higgs potential are identical for any lattice whose symmetry group is a superset of the Shastry--Sutherland wallpaper group \textit{p4g} and a subset of the square lattice wallpaper group \textit{p4mm}. To this end, we demonstrate that any term allowed on the Shastry--Sutherland lattice is also symmetry-allowed on the square lattice.

Due to invariance under $\mathcal{T}$, any such term is invariant under $P_x$ and $P_y$, and the presence of $G_x$ implies invariance under $T_y$ (see \tabref{tab:higgs_transformations}). As we can represent the remaining generators as $T_x = P_y \circ G_x$ and $R_{\pi/2} = P_x \circ \sigma_{xy}$, it is also invariant under $T_x$ and $R_{\pi/2}$. Therefore, any symmetry-allowed term on the Shastry--Sutherland lattice automatically satisfies square lattice symmetry, and the structure of the Higgs potential is thus identical to arbitrary order.

As has been previously argued on the square lattice~\cite{Song2019}, there exists a trivial monopole for the $\U(1)$ staggered flux state~\cite{Alicea-2008}, which proliferates and is expected to break the $\U(1)$ spin liquid down to either N\'eel or VBS order~\cite{SongPRX2020, Shackleton2021}. As the Shastry--Sutherland symmetry group is a subset of the one of the square lattice, this monopole still transforms trivially under the Shastry--Sutherland symmetry group and PSG, so we expect the same instability towards N\'eel and VBS order.

Similarly, we expect the fate of the $\SU(2)$ $\pi$-flux state to lead to N\'eel or one of the columnar or plaquette VBS orders~\cite{Wang_2017}. A large-$N_f$ analysis of the $\pi$-flux to $\mathbb{Z}_2$ spin liquid transition~\cite{Shackleton2021} finds enhanced N\'eel correlations at the critical point, suggesting that the $\SU(2)$ $\pi$-flux state may prefer to confine to N\'eel order, whilst the U800 spin liquid confines to a VBS state. However, we stress that the $\SU(2)$ $\pi$-flux theory on its own has no preference to either phase, and that different points in the phase diagram even in the same spin liquid phase could lead to different confined phases depending on microscopic details. Therefore, it is possible that some parts of the $\U(1)$ spin-liquid phase confine to columnar VBS order, whilst others confine to plaquette VBS order as observed numerically~\cite{Sandvik2007,Ralko-2009,Corboz2013,Xi-2023,Liu-2024,LiuESS-2024,Maity2024,Qian-2025}. 

\subsection{Symmetry-allowed couplings}
In the form of \eqnref{eq:full_lagrangian}, by ignoring the gradient terms, which have a larger scaling dimension and are thus less relevant, the critical theory is identical to the square lattice case~\cite{Shackleton2021}. As the mean-field \textit{ansatz} on the lattice might lack terms that are symmetry-allowed in the continuum theory, we now analyze all allowed fermion bilinears that respect time reversal $\mathcal{T}$, which acts as $\mathcal{T}=i\sigma^y K$, as well as lattice symmetries. Time-reversal symmetry is implemented identically on both the square and Shastry--Sutherland lattice. For the $\SU(2)$ $\pi$-flux state and U800 up to nearest-neighbor hopping, their PSG is consistent with the square lattice symmetries defined as 
\begin{equation}
\begin{split}
    T_x: (i_x,i_y) &\mapsto (i_x+1,i_y)\,,\\
    T_{y}: (i_x,i_y) &\mapsto (i_x,i_y+1)\,,\\
    P_x: (i_x,i_y) &\mapsto (-i_x,i_y)\,,\\
    P_y: (i_x,i_y) &\mapsto (i_x,-i_y)\,,\\
    R_{\pi/2}: (i_x,i_y) & \mapsto (-i_y,i_x)\,.
\end{split}
\end{equation}
The square lattice symmetry is explicitly broken for the Z3000 \textit{ansatz}, which only preserves the Shastry--Sutherland symmetry group $p4g$ as depicted in \figref{fig:lattice}(a):
\begin{equation}
\begin{split}
    T_{2x}: (i_x,i_y) &\mapsto (i_x+2,i_y)\,,\\
    T_{2y}: (i_x,i_y) &\mapsto (i_x,i_y+2)\,,\\
    G_x: (i_x,i_y) &\mapsto (i_x+1,-i_y)\,,\\
    G_y: (i_x,i_y) &\mapsto (-i_x,i_y+1)\,,\\
    \sigma_{xy}: (i_x,i_y) &\mapsto (i_y,i_x)\,,\\
    \sigma_{x\bar{y}}: (i_x,i_y) &\mapsto (-i_y+1,-i_x+1)\,,\\
    C_4: (i_x,i_y) &\mapsto (-i_y+2,i_x-1)\,.
\end{split}
\end{equation}
The projective action for the relevant spin liquids is defined in \appref{app:PSGs}. For simplicity, we set all global $\SU(2)$ matrices to the identity. 
\subsubsection{Square lattice symmetries}

Then, the gauge-dependent Majorana fermions transform nontrivially under time reversal and the square lattice space group symmetries given by the PSG:
\begin{equation}
\begin{split} \label{eq:square_lattice_symmetries}
    T_x:\X&\rightarrow \mu^x\X\,,\\ 
    T_y: \X &\rightarrow \mu^z \X\,,\\
    P_x: \X &\rightarrow  -\rho^z\mu^z \X(-x,y)\,,\\
    P_y: \X &\rightarrow \rho^x\mu^x \X(x,-y)\,,\\ 
    \mathcal{T}: \X &\rightarrow \rho^y\mu^y \X\,, \quad  i\rightarrow -i\,,\\
    R_{\pi/2}: \X&\rightarrow e^{-i\pi \rho^y/4}e^{i\pi \mu^y/4}\X(-y,x)\,.
\end{split}
\end{equation}
For the Higgs fields $\Phi_{1,2,3}^a$, we deduce the transformation properties in \tabref{tab:higgs_transformations}. By comparing with the symmetry transformations of all fermion bilinears in~\cite{Thomson_2018}, we see that there are no other symmetry-allowed terms that one could add to the Lagrangian \eqnref{eq:full_lagrangian} on the square lattice.
\subsubsection{Shastry--Sutherland symmetries}\label{sec:ss_symmetries}
The Shastry--Sutherland lattice breaks symmetries of the square lattice. Thus, we can use the transformation properties of $\X$ in \eqnref{eq:square_lattice_symmetries} to derive the action of the Shastry--Sutherland space group \textit{p4g}, which can be generated by $G_x$ and $\sigma_{xy}$~\cite{Maity2024}. These can be expressed with respect to the square lattice symmetry generators as
\begin{equation}
    G_x=P_y\circ T_x\,, \qquad \sigma_{xy} = P_x\circ R_{\pi/2}\,.
\end{equation}
On the low-energy degrees of freedom, they act as
\begin{equation}
\begin{split}
    G_{x}: \X &\rightarrow \rho^x\X(x,-y),\\
    \mathcal{T}: \X &\rightarrow \rho^y\mu^y \X\,, \quad  i\rightarrow -i\,,\\
    \sigma_{xy}: \X &\rightarrow -\rho^z\mu^ze^{-i\pi \rho^y/4}e^{i\pi \mu^y/4}\X(y,x)\,.   
\end{split}
\end{equation}
We tabulated all possible bilinears up to first-order gradient terms in \tabref{tab:SS_bilinears_1} to \tabref{tab:SS_bilinears_3}.
\begin{table}
    \centering
    \begin{tblr}{
              width = 0.49\linewidth,
              colspec = {c X[c] X[c] X[c]},
            }\hline\hline
            $T^j$ & $G_x$ & $\mathcal{T}$ & $\sigma_{xy}$\\\hline
            $\mu^y$ & $ -$& $- $ & $ +$  \\
            $\tau^b$ & $ -$ & $- $ & $ -$  \\
            $\mu^x\tau^b$ & $ -$ & $+ $  & $ -\mu^z\tau^b$ \\
            $\mu^z \tau^b$ & $ -$  & $ +$ & $ -\mu^x\tau^b$ \\\hline
        \end{tblr}
    \caption{Transformation of $\Tr[\tau^a \bar{\mathcal{X}}T^j \X]$ under the Shastry--Sutherland symmetries. $T^j=\lbrace \mu^y,\tau^b, \mu^x\tau^b,\mu^z\tau^b\rbrace$ are the $10$ generators of $\SO(5)$.}
    \label{tab:SS_bilinears_1}
\end{table}
\begin{table}
    \centering
    \begin{tblr}{
          width = 0.49\linewidth,
          colspec = {c X[c] X[c] X[c]},
          row{3} = {bg=colone},
          row{4} = {bg=coltwo},
          row{6} = {bg=colone},
          row{7} = {bg=coltwo},
        }\hline\hline
        $\Gamma^j \gamma^\mu$ & $G_x$ & $\mathcal{T}$ & $\sigma_{xy}$\\\hline
        $\mu^x\gamma^0$ & $ +$& $+ $ & $ \mu^z\gamma^0$  \\
        $\mu^x\gamma^x$ & $ +$& $ -$ & $ \mu^z \gamma^y$  \\
        $\mu^x\gamma^y$ & $ -$& $ -$ & $ \mu^z\gamma^x$  \\\hline
        $\mu^z\gamma^0$ & $ +$& $ +$ & $ \mu^x\gamma^0$  \\
        $\mu^z\gamma^x$ & $ +$& $ -$ & $ \mu^x\gamma^y$  \\
        $\mu^z\gamma^y$ & $ -$& $ -$ & $ \mu^x\gamma^x$  \\\hline
        $\mu^y \tau^b\gamma^0$ & $ +$& $ -$ & $ -$ \\
        $\mu^y \tau^b\gamma^x$ & $ +$& $ +$ & $ -\mu^y\tau^b\gamma^y$  \\
        $\mu^y \tau^b\gamma^y$ & $ -$& $ +$ & $ -\mu^y\tau^b\gamma^x$  \\\hline
    \end{tblr}
    \caption{Transformation of $\Tr[\tau^a \bar{\mathcal{X}}\Gamma^j\gamma^\mu \X]$ under the Shastry--Sutherland symmetries. $\Gamma^j=\lbrace \mu^x,\mu^z, \mu^y\tau^b\rbrace$ transforms as a vector under $\SO(5)$.}
    \label{tab:SS_bilinears_1_2}
\end{table}
\begin{table*}
    \centering
    \begin{minipage}[t]{0.49\linewidth}
        \centering
        \begin{tblr}{
              width = \linewidth,
              colspec = {c X[c] X[c] X[c]},
              row{3} = {bg=colone},
              row{4} = {bg=coltwo},
              row{5} = {bg=colone},
              row{7} = {bg=colthree},
              row{8} = {bg=coltwo},
              row{9} = {bg=colthree},
            }\hline\hline
            $\mu^y\gamma^\mu i\partial_\nu$ & $G_x$ & $\mathcal{T}$ & $\sigma_{xy}$\\\hline
            $\mu^y\gamma^0 i\partial_0$ & $+ $& $+ $ & $-$  \\
            $\mu^y\gamma^0 i\partial_x$ & $+ $& $- $ & $-\mu^y\gamma^0i\partial_y $  \\
            $\mu^y\gamma^0 i\partial_y$ & $- $& $- $ & $-\mu^y\gamma^0i\partial_x $  \\\hline
            $\mu^y\gamma^x i\partial_0$ & $+ $& $- $ & $-\mu^y\gamma^yi\partial_0 $  \\
            $\mu^y\gamma^x i\partial_x$ & $+ $& $+ $ & $-\mu^y\gamma^yi\partial_y $  \\
            $\mu^y\gamma^x i\partial_y$ & $- $& $+ $ & $-\mu^y\gamma^yi\partial_x $  \\\hline
            $\mu^y\gamma^y i\partial_0$ & $- $& $- $ & $-\mu^y\gamma^xi\partial_0 $  \\
            $\mu^y\gamma^y i\partial_x$ & $- $& $+ $ & $-\mu^y\gamma^xi\partial_y $  \\
            $\mu^y\gamma^y i\partial_y$ & $+ $& $+ $ & $-\mu^y\gamma^xi\partial_x $  \\\hline
        \end{tblr}
    \end{minipage}
    \hfill
    \begin{minipage}[t]{0.49\linewidth}
        \centering
        \begin{tblr}{
              width = \linewidth,
              colspec = {c X[c] X[c] X[c]},
              row{2} = {bg=colthree},
              row{3} = {bg=coltwo},
              row{4} = {bg=colone},
            }\hline\hline
            $\mu^{0,x,z}i\partial_\mu$ & $G_x$ & $\mathcal{T}$ & $\sigma_{xy}$\\ \hline 
            $i\partial_0$ & $ -$& $ +$ & $- $  \\
            $i\partial_x$ & $ -$& $ -$ & $-i\partial_y $  \\
            $i\partial_y$ & $ +$& $ -$ & $-i\partial_x $  \\\hline
            $\mu^xi\partial_0$ & $ -$ & $- $ & $-\mu^zi\partial_0$ \\
            $\mu^xi\partial_x$ & $ -$ & $ +$ & $-\mu^zi\partial_y$ \\
            $\mu^xi\partial_y$ & $ +$ & $ +$ & $-\mu^zi\partial_x$ \\\hline
            $\mu^zi\partial_0$ & $ -$ & $- $ & $-\mu^xi\partial_0$ \\
            $\mu^zi\partial_x$ & $ -$ & $ +$ & $-\mu^xi\partial_y$ \\
            $\mu^zi\partial_y$ & $ +$ & $ +$ & $-\mu^xi\partial_x$ \\\hline
        \end{tblr}
    \end{minipage}
    \caption{Transformation of $\Tr[\tau^a \bar{\X}i\partial_\mu\X],\Tr[\tau^a\bar{\X}\Gamma^ji\partial_\mu \X],$ and $\Tr[\tau^a \bar{\X}T^j\gamma^\mu i \partial_\nu \X]$ under the Shastry--Sutherland symmetries. We only consider terms that do \textit{not} transform under spin~\cite{Thomson_2018}.}
    \label{tab:SS_bilinears_3}
\end{table*}
We color code all fermion bilinears, which do not transform under spin and are consistent with the transformation properties of the Higgs fields in \tabref{tab:higgs_transformations}. For $\Phi_3$, there are only additional gradient terms, whilst for $\Phi_{1/2}$, we have additional zeroth-order bilinears and gradient terms. As the gradient terms are less relevant, we can ignore them. The zeroth-order bilinears change the coupling between the Higgs fields and Dirac fermions but as we will show in \secref{sec:rg}, this does not alter the scaling dimensions of the quantum field theory.

\subsubsection[\texorpdfstring{$\SU(2)$ $\pi$-flux state}{SU(2) pi-flux state}]{\texorpdfstring{$\bm{\SU(2)}$ $\bm{\pi}$-flux state}{SU(2) pi-flux state}}
This phase corresponds to no Higgs condensate $\langle \Phi_{1,2,3}^a\rangle =0$ and possesses an $\SU(2)$ gauge symmetry as well as an emergent $\SO(5)$ symmetry as discussed in \secref{sec:su2_continuum}. The theory is believed to confine to the N\'eel or VBS phase~\cite{Wang_2017,WangPRL2021}.
\subsubsection{U800 staggered flux state}
The staggered-flux state corresponds to condensing $\langle \Phi_3^a\rangle \propto \delta\theta \delta_{a,z}$. This theory has a single gauge boson, which, as discussed in \secref{sec:maj_higgs} and Ref.~\cite{Shackleton2021}, we believe likely confines to a VBS state, but possibly also to the N\'eel phase~\cite{Alicea-2008,SongPRX2020}.
\subsubsection{Z3000 state}
The Z3000 state corresponds to condensing $\langle \Phi_{1,2}^a\rangle\neq 0$. Due to the symmetry transformations between the two fields, there is only a single tuning parameter $s$ to condense them, and the existence of the term $w$ implies that also $\langle \Phi_3^a\rangle\propto w\epsilon_{abc}\langle \Phi_{1}^b\rangle\langle \Phi_{2}^c\rangle$ condenses.

As was shown in a recent PSG study~\cite{Maity2024}, this spin liquid is related to the Z2A$zz$13 spin liquid on the square lattice~\cite{Wen2002}. This is consistent with our conclusion that up to less relevant gradient terms the continuum theory is identical to the one on the square lattice~\cite{Shackleton2021}. Similarly, the structure of the symmetry-allowed Higgs potential is identical for the two systems to arbitrary order in the Higgs fields.
\section{\texorpdfstring{Renormalization study of the critical $\bm{\SU(2)}$ gauge theory}{Renormalization study of the critical SU(2) gauge theory}}
\label{sec:rg}

The stability of the emergent $\SO(5)$ symmetry at the deconfined quantum critical point
remains a subject of active debate. While the fermionic $\SU(2)$ $\pi$-flux theory exhibits an enlarged $\SO(5)$ symmetry that unifies the N\'eel and valence-bond-solid order parameters~\cite{Senthil2004a,Senthil2004b,Nahum2015,Wang_2017}, a growing body of analytical and numerical work indicates that this fixed point is generically unstable in the strict infrared due to symmetry-allowed perturbations, most notably the Yukawa coupling between Dirac spinons and Higgs fields~\cite{Wang_2017,WangPRL2021,Bowen_2020}.
Importantly, this instability does not preclude the appearance of robust $\SO(5)$-symmetric scaling over extended intermediate length scales. Instead, it is now widely understood that the $\SO(5)$ fixed point acts as a pseudocritical or ``walking'' fixed point, controlling a broad pre-asymptotic regime before renormalization-group flows ultimately run away at the lowest energies~\cite{Shao2016Science,Nahum2015PRX,Zhou_2024,Takahashi2024}.
This picture reconciles large-$N$ field-theoretic analyses with state-of-the-art Monte
Carlo simulations, which consistently observe approximate $\SO(5)$ symmetry,
emergent operator degeneracies, and universal scaling over many decades despite an
eventual departure from true conformality.
In this sense, the $\SO(5)$ theory provides the appropriate organizing framework for
both the square-lattice and Shastry--Sutherland antiferromagnets, even though it does
not represent a stable infrared fixed point. To assess the ultimate fate of this fixed point, we reformulate the theory in terms of $N_f=2$ flavors of complex Dirac fermions, generalizing the model to include $v=1,\dots,N_f$ valleys and a Higgs flavor index $\alpha=1,\dots,N_b/2$ to facilitate a controlled expansion. 

The Dirac fermions can be expressed with respect to the Majorana matrix operator as
\begin{equation}\label{eq:dirac_fermions}
    \psi_{a,m_x,v} = i \tau_{ab}^y[\X_{m_x,v}]_{1,b}\,.
\end{equation}
Introducing the $\SU(2)$ gauge field $A_\mu^a$ via minimal coupling leads to the effective Lagrangian:
\begin{equation}
\begin{split}
    \mathcal{L}&=\mathcal{L}_\psi + \mathcal{L}_\Phi + \mathcal{L}_{\Phi\psi}\,,\\
    \mathcal{L}_\psi&=i\bar{\psi}_v\gamma^\mu(\partial_\mu-iA_\mu^a\tau^a)\psi_v\,,\\
    \mathcal{L}_{\Phi}&=\frac{1}{2g}[(\partial_x\Phi_1^a+2\epsilon_{abc}A_x^b\Phi_1^c)^2+(\partial_y\Phi_2^a+2\epsilon_{abc}A_y^b\Phi_2^c)^2]+\dots\,,\\
    \mathcal{L}_{\Phi\psi}&=y\sum_{\alpha}(\Phi_{1\alpha}^a\bar{\psi}_v\mu^z\gamma^x\tau^a\psi_v+\Phi_{2\alpha}^a\bar{\psi}_v\mu^x\gamma^y\tau^a\psi_v)\,.
\end{split}
\end{equation}
This field theory is identical to the square lattice case~\cite{Shackleton2021,Shackleton:2022zzm}. Therefore, we can conclude that also on the level of the quantum field theory the spin liquid and the associated transitions towards the N\'eel and VBS phase on the square lattice can be connected to the ones on the Shastry--Sutherland antiferromagnet. Unlike previous work in the $y\rightarrow \infty$ limit~\cite{Shackleton2021}, which breaks the emergent $\SO(5)$ symmetry and also Lorentz invariance, we will study the $\SO(5)$ critical point by considering $y$ as a perturbation and assessing the effect of the Yukawa coupling on the stability of the phase in this perturbative expansion. We want to emphasize that the following field-theoretic analysis applies to both the square as well as the Shastry--Sutherland Heisenberg antiferromagnet.

The analysis in this section focuses on the $\SO(5)$-symmetric theory at $y=0$, which is expected to be exactly conformal. The influence of a non-zero $y$ is addressed in \secref{sec:yukawa}.

As all symmetry-allowed terms should be present in the Lagrangian, we will add a quadratic minimally coupled kinetic term for each boson $\Phi^a_{s,\alpha},\alpha=1,\dots,N_b/2,s=1,2,a=1,2,3$ in all space-time directions. In general, these terms could have different coupling constants, but we will assume they coincide in the following. This leads to the following theory:
\begin{equation}
\begin{split}
    \mathcal{L}_{\psi}&=i\bar{\psi}_v\gamma^\mu(\partial_\mu-iA_\mu^a\tau^a)\psi_v\,,\\
    \mathcal{L}_{\Phi}&=\frac{1}{2g}[(\delta_{ac}\partial_\mu+2\epsilon_{abc}A_\mu^b)\Phi_{s\alpha}^c]^2\,,\\ 
    \mathcal{L}_{\Phi\psi}&=y\sum_{s,\alpha}\Phi_{s\alpha}^a\bar{\psi}_vX^s\tau^a\psi_v\,,
\end{split}
\end{equation}
where $X^s=(\mu^z\gamma^x,\mu^x\gamma^y)$ and $s=1,2$.
To resolve the constraint $\sum_{s,\alpha} (\Phi_{s\alpha}^a)^2=1$, we add a real-valued field $\lambda$ as a Lagrange multiplier, which allows us to write the bosonic action by rescaling $\Phi_{s\alpha}^a \rightarrow \sqrt{\frac{g}{3N_b}}\Phi_{s\alpha}^a$ as
\begin{equation}
    \mathcal{L}_{\Phi}=\frac{1}{2g}\left[((\delta_{ac}\partial_\mu+2\epsilon_{abc}A_\mu^b)\Phi^c_{s\alpha})^2+i\lambda\left((\Phi_{s\alpha}^a)^2-\frac{3N_b}{g}\right)\right]\,.
\end{equation}
We now follow~\cite{Polyakov_1987} to obtain the saddle-point equation in the large-$N_b$ limit: one can first eliminate $A_\mu$ by minimizing $S$ with respect to $A_\mu$. Then, we integrate out the bosons leading to
\begin{equation}
    S_b=-\frac{3N_b}{2g}\int \dd^3r\, i\lambda+\frac{3N_b}{2}\log\det|-\partial_\mu^2+i\lambda|\,.
\end{equation}

For large $N_b$, the only relevant contributions to the path integral are at the saddle point, which we can obtain by varying the action with respect to $\lambda$:
\begin{equation}
    \frac{3N_bi}{2g}=\frac{3N_b}{2}\frac{\delta}{\delta \lambda}\log\det|-\partial_\mu^2+i\lambda|=\frac{3N_bi}{2}G_\Phi(x,x)\,,
\end{equation}
where we have introduced the Green's function
\begin{equation}
    G_\Phi=(-\partial_\mu^2+i\lambda)^{-1}\,.
\end{equation}
We can solve the saddle point equation by going to momentum space with $G_\Phi^{-1}=k^2+i\lambda$:
\begin{equation}
    \frac{1}{g}=\int \frac{\dd^3 k}{8\pi^3}\frac{1}{k^2+i\lambda}\,.
\end{equation}
This is solved by $r=i\lambda_0$ with $G_\Phi(k)=(k^2+r)^{-1}$. For large $N_b$ this theory becomes critical when $r=0$~\cite{Kaul_2008}.

We now obtain the effective gauge theory for $\lambda$ and $A_{\mu}^{a}$ at $T=0$ by expanding around the saddle point leading to a correction to the action
\begin{equation}
    \mathcal{F}=\frac{1}{2}\int\frac{\dd^3 p}{(2\pi)^3}\left(\Pi_\lambda\lambda^2+\Pi_{A}\left(\delta_{\mu\nu}-\frac{p_\mu p_\nu}{p^2}\right)A_\mu^a A_\nu^a\right)\,,
\end{equation}
where 
\begin{equation}
\begin{split}\label{eq:polarization_lambda_A}
    \Pi_\lambda(p,r)&=\frac{3N_b}{8\pi p}\arctan \frac{p}{2\sqrt{r}}\,,\\
    \Pi_A(p,r)&=2N_f\frac{p}{16} + 4N_b\left[\frac{p^2+4r}{8\pi p}\arctan \frac{p}{2\sqrt{r}}-\frac{\sqrt{r}}{4\pi}\right]\,.
\end{split}
\end{equation}
\begin{figure}
    \centering
    \includegraphics[width=0.4\linewidth]{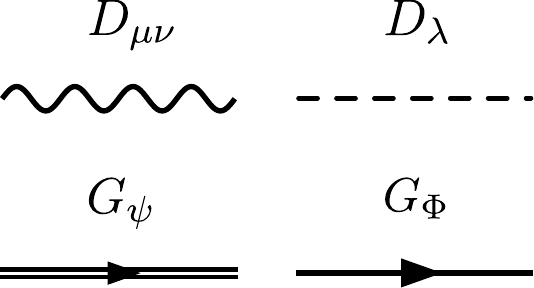}
    \caption{Definition of diagrammatic symbols for the propagators $D_{\mu\nu},D_\lambda,G_\psi,$ and $G_\Phi$.}
    \label{fig:propagator_symbols}
\end{figure}
Details are given in \appref{app:saddle_point_expansion}. The propagators are shown in \figref{fig:propagator_symbols} and can be obtained by imposing the gauge $k_\mu A_\mu=1-\zeta$:
\begin{equation}
\begin{split}
    D_\lambda &=\langle \lambda \lambda \rangle = \frac{1}{\Pi_\lambda}\,,\\
    D_{\mu\nu}^{ab} &=\langle A_\mu^aA_\nu^b\rangle = \frac{\delta_{ab}}{\Pi_A}\left(\delta_{\mu\nu}-\zeta \frac{p_\mu p_\nu}{p^2}\right)\,,\\
    G_\Phi^{ab} &= \frac{\delta_{ab}}{k^2+r}\,,\\
    G_\psi &= \frac{\slashed{k}}{k^2}\,.
\end{split}
\end{equation}
As we are interested in the critical point, we will set $r=0$ in the following, so the polarizations of \eqnref{eq:polarization_lambda_A} simplify to
\begin{equation}
    \Pi_\lambda(p) = \frac{3N_b}{16 p}\,, \quad \Pi_A(p)=(2N_f+4N_b)\frac{p}{16}\,.
\end{equation}

\subsection{Fermion self-energy}
The scaling dimension of $\psi$ is defined as
\begin{equation}
    \dim[\psi]=\frac{D-1+\eta_\psi}{2}\,
\end{equation}
with $D=3$ and the anomalous dimension $\eta_\psi$. We can evaluate $\eta_\psi$ by calculating the lowest order self-energy graph $\Sigma(k)$ due to the gauge field, see \figref{fig:fermion_self_energy_correction}, and picking up the logarithmically divergent $\slashed{k}\log k$ part:
\begin{figure}[t]
    \centering
    \includegraphics[width=0.4\linewidth]{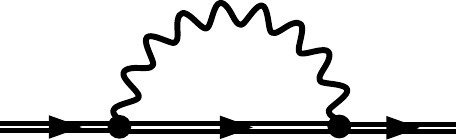}
    \caption{Self-energy correction to the fermion propagator due to the gauge field.}
    \label{fig:fermion_self_energy_correction}
\end{figure}
\begin{equation}
\begin{split}\label{eq:fermion_selfenergy}
    \Sigma_\psi(k)&=\tau^a\tau^a\int \frac{\dd^3q}{(2\pi)^3}\gamma_\mu G_\psi(k+q) \gamma_\nu D_{\mu\nu}(-q)\rightarrow\frac{8}{(2N_f+4N_b)\pi^2}\left(1-3\zeta\right)\slashed{k}\log k\,.
\end{split}
\end{equation}
Details are given in \appref{app:useful_integrals}. This gives the gauge-dependent anomalous dimension 
\begin{equation}
    \eta_\psi=\frac{8(1-3\zeta)}{(2N_f+4N_b)\pi^2}
\end{equation}
and the overall fermion scaling dimension
\begin{equation}
    \dim[\psi]=1+\frac{\eta_\psi}{2}=1+\frac{4}{(2N_f+4N_b)\pi^2}(1-3\zeta)\,.
\end{equation}

\subsection{Boson self-energy}
\begin{figure}[t]
    \centering
    \includegraphics[width=0.8\linewidth]{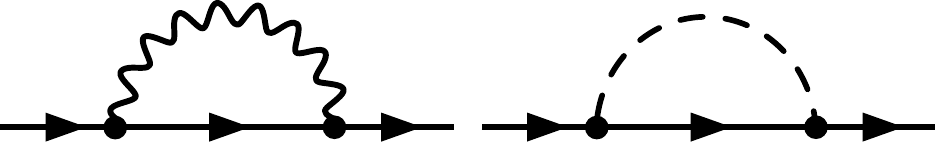}
    \caption{Self-energy corrections to the Higgs propagator due to the gauge field and the Lagrange multiplier field.}
    \label{fig:boson_self_energy_diagrams}
\end{figure}
The scaling dimension of the Higgs fields $\Phi$ is similarly given by
\begin{equation}
    \dim[\Phi] = \frac{D-2+\eta_\Phi}{2}\,
\end{equation}
with anomalous dimension $\eta_\Phi$. We can obtain that to $\mathcal{O}(1/N)$ order by extracting the $k^2\log k$ divergence of the contributing diagrams in \figref{fig:boson_self_energy_diagrams}.  We begin with the corrections due to the gauge field:
\begin{equation}
\begin{split}\label{eq:I_A1}
    I_{A;1}&=8\int \frac{\dd^3 q}{(2\pi)^3}G_\Phi(k+q)D_{\mu\nu}(-q)(2k+q)_\mu (2k+q)_\nu \\
    &\rightarrow -\frac{8\cdot 4}{(2N_f+4N_b)\pi^2}\left(\frac{10}{3}+2\zeta\right)k^2 \log k\,.
\end{split}
\end{equation}
The factor of $8$ is due to the gauge indices
\begin{equation}
    2\epsilon_{abc}\cdot 2\epsilon_{fbc}=8\delta_{af}\,,
\end{equation}
where $a$ and $f$ are the ingoing and outgoing gauge indices of the bosons. 

For the Lagrange multiplier fluctuations, we obtain
\begin{equation}
\begin{split}\label{eq:I_lambda1}
    I_{\lambda;1} &= i^2 \int \frac{\dd^3 q}{(2\pi)^3}G_\Phi(k+q)D_\lambda(-q) \rightarrow \frac{8}{9N_b\pi^2}k^2\log k\,.
\end{split}
\end{equation}
Therefore, we can conclude
\begin{equation}
    \eta_{\Phi}=-\frac{32}{(2N_f+4N_b)\pi^2}\left(\frac{10}{3}+2\zeta\right)+\frac{8}{9N_b\pi^2}\,.
\end{equation}
\subsection[{Renormalization of the \texorpdfstring{${\SO(5)}$}{SO(5)} order parameter}]{Renormalization of the \texorpdfstring{$\bm{\SO(5)}$}{SO(5)} order parameter}
\begin{figure}[t]
    \centering
    \includegraphics[width=0.6\linewidth]{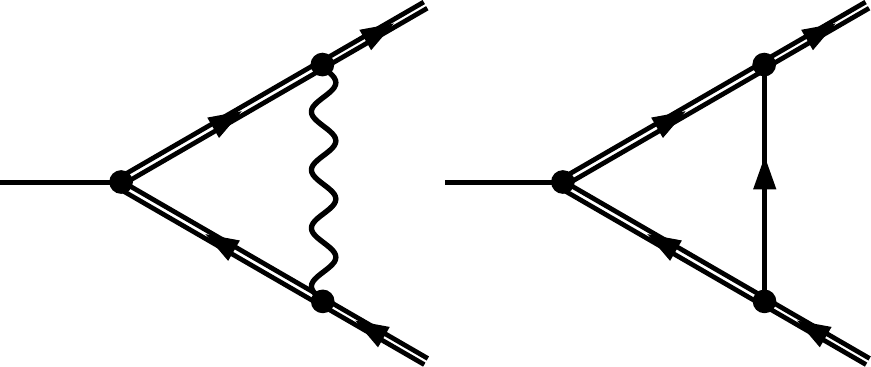}
    \caption{Vertex corrections to the $\SO(5)$ order parameter due to the gauge field and the Yukawa coupling of the fermions to the Higgs fields.}
    \label{fig:so5_renormalization_diagrams}
\end{figure}
As discussed in \secref{sec:su2_continuum}, the $\pi$-flux state possesses an emergent $\SO(5)$ symmetry combining the N\'eel and VBS order parameters into the $5$-vector 
\begin{equation}
    i\Tr[\bar{\X}\Gamma^j\X]
\end{equation}
with 
\begin{equation}
    \Gamma^j=\lbrace \mu^x,\mu^z,\mu^y\tau^a\rbrace\,.
\end{equation}
The first two components correspond to the VBS order, whilst the last three correspond to the N\'eel order. We begin by expressing the VBS order parameter with respect to the Dirac fermions in \eqnref{eq:dirac_fermions}:
\begin{equation}
    V^i=(\bar{\psi} \mu^x \psi, \bar{\psi}\mu^z \psi)\,.
\end{equation}
For the N\'eel order, we will only consider the $\hat{z}$-component. The renormalization of the other components is identical due to $\SU(2)$ symmetry:
\begin{equation}
    N^z=\bar{\psi}\mu^y\psi\,.
\end{equation}

We therefore have to study the anomalous dimensions of the bilinears $n^i=\bar{\psi}\mu^i\psi$ with $i=x,y,z$, which we do by introducing a source field $J^i$ and calculating the vertex corrections as shown in \figref{fig:so5_renormalization_diagrams}. We begin with the corrections due to the gauge field:
\begin{equation}
\begin{split}
    I_{A;2}&=3 \int \frac{\dd^3 q}{(2\pi)^3} \gamma_\mu G_\psi(k_1+q) G_\psi(k_2+q) \gamma_\nu D_{\mu\nu}(-q)\\
    &\rightarrow \frac{3\cdot 8\mu^i}{(2N_f+4N_b)\pi^2}\left(\zeta-3\right)\log k\,,
\end{split}
\end{equation}
where we restricted ourselves to the case $k_1=k_2\equiv k$.

The corrections due to the Higgs field do not have a logarithmic divergence, so they can be ignored. We can now collect the vertex contribution as well as the self-energy correction to the fermion bilinear:
\begin{equation}
\begin{split}
    \dim[\bar{\psi}\mu^i\psi]&=2\dim[\psi] + \eta_{\mathrm{vrtx}} =2-\frac{64}{(2N_f+4N_b)\pi^2}\,.\label{eq:dim_so5}
\end{split}
\end{equation}
Reassuringly, this result is gauge-invariant. Compared to, e.g., Ref.~\cite{Kaul_2008}, the corrections to the scaling dimension are a factor of $3$ larger due to having an $\SU(2)$ instead of $\U(1)$ gauge field. Both the VBS and N\'eel order parameters receive the same correction.

We can use this to obtain the correction to the scaling of the susceptibility
\begin{equation}
    \chi(k)=\langle\bar{\psi}\Gamma^i\psi(k)\bar{\psi}\Gamma^j\psi(-k)\rangle\sim |k|^{1-\frac{128}{(2N_f+4N_b)\pi^2}}\sim |k|^{-0.081}\,
\end{equation}
for the physically relevant values $N_f=N_b=2$. 

The exponent is negative, so the susceptibility diverges as $k \to 0$, but only weakly, a consequence of the large anomalous dimension $\eta$ associated with deconfined quantum criticality~\cite{Senthil2004a}. In real space, the susceptibility follows the dual scaling:
\begin{equation}
    \chi(r) \sim |r|^{-4 + \frac{128}{(2N_f+4N_b)\pi^2}} \approx |r|^{-2.919}\,.
\label{eq:chi_r}
\end{equation}
In the physical case $N_f=N_b=2$, Eq.~\eqref{eq:dim_so5} gives $\dim[\bar{\psi}\mu^i\psi]\approx 1.46$ for the $\SO(5)$ order parameter, i.e., a large anomalous dimension $\eta_{\SO(5)}=2\dim[\bar{\psi}\mu^i\psi]-1\approx 1.92$ in the convention $\chi(r)\sim|r|^{-(1+\eta)}$. This is in sharp contrast to the $\mathrm{O}(3)$ universality class, whose anomalous dimension is small~\cite{Sandvik_2020, Nahum2015PRX}. Such large anomalous dimensions are a hallmark of fermionic deconfined quantum criticality and reflect the fractionalized nature of the critical degrees of freedom.

Large anomalous dimensions have also been reported in numerical studies of $J$\textendash$Q$ models, which find $\eta \approx 0.3$--$0.35$ for the N\'eel--VBS order parameter~\cite{Kaul}. Our result is obtained in a large-$N$ expansion, and at $N_f=N_b=2$ the $\mathcal{O}(1/N)$ value of $\eta_{\SO(5)}$ is considerably larger than these estimates, so it should not be read quantitatively. The common feature, an anomalous dimension far above that of the $\mathrm{O}(3)$ class, supports the interpretation of both systems as exhibiting pseudocritical $\SO(5)$-symmetric behavior governed by fractionalized degrees of freedom.

\subsection{Correlation length exponent}
\begin{figure}[t]
    \centering
    \includegraphics[width=0.9\linewidth]{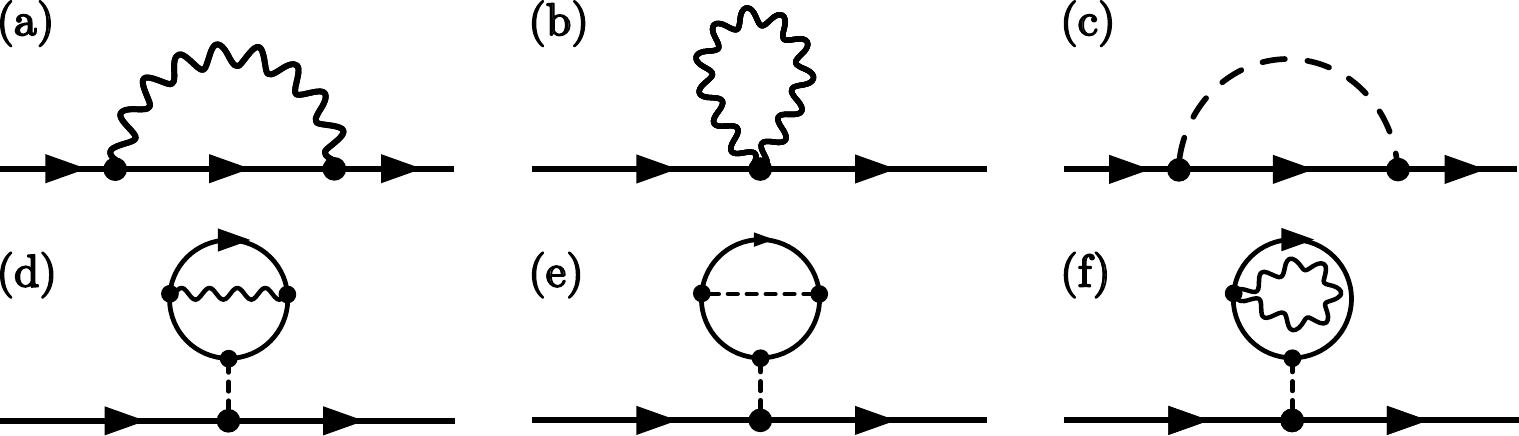}
    \caption{Self-energy diagrams of $\Phi$ at order $\mathcal{O}(1/N)$. At the critical point, only (a) and (c) contribute. These diagrams are used to calculate $\gamma_\Phi$ and, via the scaling relation $\gamma_\Phi=\nu(2-\eta_\Phi)$, the correlation-length exponent.}
    \label{fig:correlation_length_contributions}
\end{figure}

We compute the correlation length exponent $\nu$ defined via
\begin{equation}
    \xi \propto (g-g_c)^{-\nu}\,,
\end{equation}
following~\cite{Kaul_2008}. The critical exponent is related to
\begin{equation}
    \nu = \frac{\gamma_\Phi}{2-\eta_\Phi}\,,
\end{equation}
where $\eta_\Phi$ is the anomalous dimension of the Higgs fields and $\gamma_\Phi$ is defined as the scaling of the mass to $0$ at the critical point:
\begin{equation}
    G_\Phi^{-1}(k=0)\sim (g-g_c)^{\gamma_\Phi}\,.
\end{equation}
Analogously to~\cite{Kaul_2008}, we define
\begin{equation}
    \frac{1}{g_c}-\frac{1}{g}=\frac{\sqrt{r_g}}{4\pi}\,.
\end{equation}
As shown in~\cite{Kaul_2008}, $\gamma_\Phi$ is related to the coefficient $\alpha$ of the logarithmic divergence $r_g\log(r_g)$ of 
$\Sigma(0,r)-\frac{\Pi_\lambda(0,0)}{\Pi_\lambda(0,r)}\Sigma(0,0)$ via
\begin{equation}
    \gamma_\Phi = 2(1-\alpha)\,.
\end{equation}
We will thus calculate $\alpha$ in the following. The contributing diagrams are shown in \figref{fig:correlation_length_contributions}, and their contributions to the self-energy are:
\begin{equation}
\begin{split}
    \Sigma^{(a)} &= I_{A;1}\,,\\ 
    \Sigma^{(b)} &= 8 \sum_{\mu,\nu}\int \frac{\dd^3 q}{(2\pi)^3}\frac{1}{\Pi_A(q)}\left(\delta_{\mu\nu}-\zeta \frac{q_\mu q_\nu}{q^2}\right)\,,\\
    \Sigma^{(c)} &= I_{\lambda;1}\,,\\
    \Sigma^{(d)} &= \frac{i^2}{\Pi_\lambda(0,r)}\int \frac{\dd^3 q}{(2\pi)^3}\frac{3N_b}{2}G_\Phi(q)^2I_{A;1}(q)\,,\\
    \Sigma^{(e)} &= \frac{i^2}{\Pi_\lambda(0,r)}\int \frac{\dd^3 q}{(2\pi)^3}\frac{3N_b}{2}G_\Phi(q)^2I_{\lambda;1}(q)\,,\\
    \Sigma^{(f)} &= \frac{i^2\Sigma^{(b)}}{\Pi_{\lambda}(0,r)}\int \frac{\dd^3 q}{(2\pi)^3}\frac{3N_b}{2}G_\Phi(q)^2 = - \Sigma^{(b)}\,.\\
\end{split}
\end{equation}
For $\Sigma = \Sigma^{(c)}+\Sigma^{(e)}$, we get the contribution
\begin{equation}
    \alpha_{\lambda}=\frac{4}{N_b\pi^2}\,.
\end{equation}
We now look at the gauge contributions $\Sigma=\Sigma^{(a)}+\Sigma^{(d)}$: All steps in~\cite{Kaul_2008} up to Eq.~(31) still hold true for our case with an additional factor of $8$ leading to
\begin{equation}
\begin{split}
    \Sigma(0,r)-\frac{\Pi_\lambda(0,0)}{\Pi_\lambda(0,r)}\Sigma(0,0)=&8\int \frac{q^2 \dd q}{2\pi^2}\Bigg[ \frac{1-\zeta}{\Pi_A(q,r_g)}\frac{q^2}{q^2+r_g}\\ 
    &-\frac{3N_b}{2\Pi_{\lambda}(0,r_g)}\frac{I_A(q,r_g)}{\Pi_A(q,r_g)}+\frac{3N_b}{2\Pi_\lambda(0,r_g)}\frac{1}{4q\Pi_A(q,0)}\Bigg]\,,
\end{split}
\end{equation}
where $I_A$ is defined as in~\cite{Kaul_2008} and evaluates to
\begin{equation}
    I_A(q,r_g) \approx (1-\zeta) \frac{2\Pi_\lambda(0,r_g)}{3N_b}+\frac{1}{4q}-\frac{\sqrt{r_g}}{\pi q^2}\,.
\end{equation}
By expanding up to $1/q^3$ for large $q$, we obtain the logarithmic divergence
\begin{equation}
\begin{split}
    \alpha_A=-\frac{8\cdot4}{\pi^2(2N_f+4N_b)}\left(\zeta+\frac{14N_f-36N_b}{2N_f+4N_b}\right)\,.
\end{split}
\end{equation}
Overall, we have $\alpha=\alpha_\lambda+\alpha_A$ and 
\begin{equation}
    \gamma_\Phi=2(1-\alpha)=2-\frac{8}{N_b\pi^2}+\frac{64}{\pi^2}\left(\frac{14N_f-36N_b}{(2N_f+4N_b)^2}+\frac{\zeta}{2N_f+4N_b}\right)\,.
\end{equation}
We can now calculate the correlation length exponent:
\begin{align}
    \nu&=\frac{\gamma_\Phi}{2-\eta_\Phi}\approx \frac{\gamma_\Phi}{2}\left(1+\frac{\eta_\Phi}{2}\right) \nonumber \\
    &=\left[1-\frac{4}{N_b\pi^2}+\frac{32}{\pi^2}\left(\frac{14N_f-36N_b}{(2N_f+4N_b)^2}+\frac{\zeta}{2N_f+4N_b}\right)\right]\left[1+\frac{4}{9N_b\pi^2}-\frac{16}{(2N_f+4N_b)\pi^2}\left(\frac{10}{3}+2\zeta\right)\right] \nonumber \\
    &\approx 1 - \frac{32}{9N_b\pi^2} + \frac{32}{\pi^2}\frac{14N_f-36N_b}{(2N_f+4N_b)^2}-\frac{160}{3(2N_f+4N_b)\pi^2}\,.
    \label{eq:nures}
\end{align}
This result is, as expected, gauge-independent. For the physically relevant $N_f=N_b=2$, the $\mathcal{O}(1/N)$ correction exceeds the leading term (the linearized value is negative), so Eq.~\eqref{eq:nures} does not provide a quantitative estimate of $\nu$ at the physical point.

\subsection{Renormalization of the Yukawa coupling}
\label{sec:yukawa}

\begin{figure}[t]
    \centering
    \includegraphics[width=0.5\linewidth]{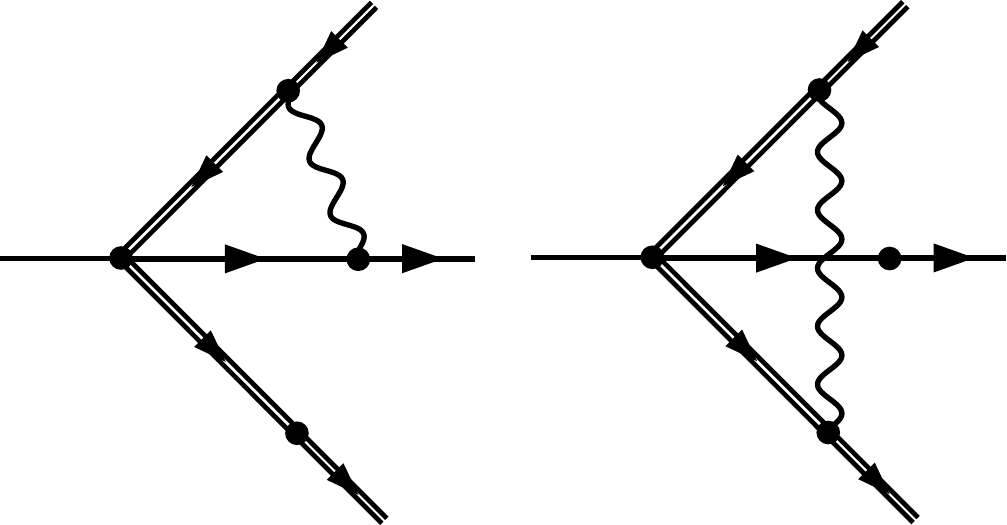}
    \caption{Vertex corrections to the Yukawa coupling. The first diagram contributes twice, by coupling to the other fermion line.}
    \label{fig:yukawa_vertex_corrections}
\end{figure}

Finally, we investigate the stability of the $\SO(5)$ critical point by calculating the vertex corrections to the scaling dimension of the Yukawa coupling $y$, which breaks both the $\SO(5)$ symmetry and Lorentz invariance. To this end, we couple the Yukawa terms to a source field and evaluate the vertex corrections mediated by the gauge field. The contributing diagrams are depicted in \figref{fig:yukawa_vertex_corrections}. Notably, the first diagram contributes twice due to the specific coupling between the Dirac fermions and the boson.

Our objective is to determine the scaling dimensions of the Yukawa coupling for different matrices $X^s$ acting on the Dirac and valley degrees of freedom. Furthermore, we aim to assess the relative relevance of other symmetry-allowed zeroth-order bilinears that transform as spin singlets (tabulated in \tabref{tab:SS_bilinears_1} and \tabref{tab:SS_bilinears_1_2}). We restrict our attention to spin singlets because terms transforming as vectors cannot form a symmetric spin liquid~\cite{Thomson_2018}. Consequently, we consider a generic Yukawa coupling to a Higgs field $\Phi$ of the form
\begin{equation}
    y\sum_{\alpha}\Phi_\alpha^a \bar{\psi}_v\tau^aX\psi_v\,.
\end{equation}

The scaling dimension of the corresponding operator is given by
\begin{equation}
\begin{split}
    \dim[\Phi^a \bar{\psi}\tau^a X\psi]&=2\dim[\psi]+\dim[\Phi]+\eta_{\mathrm{vrtx}}\,,
\end{split}
\end{equation}
which yields the scaling dimension for the coupling constant:
\begin{equation}
    \dim[y]=3-(2\dim[\psi]+\dim[\Phi]+\eta_{\mathrm{vrtx}})\,.
\end{equation}

We evaluate the first diagram by setting all external momenta to zero, as we are solely interested in extracting the logarithmic divergence:
\begin{equation}\label{eq:yukawa_renormalization_1}
\begin{split}
    I_{\Phi\psi} &= (-i\tau^b)\tau^c\int \frac{\dd^3 q}{(2\pi)^3}\gamma_\mu G_\psi(q) X G_\Phi(q)q_\nu (2\epsilon_{abc})D_{\mu\nu}(-q) \\
    &=\frac{4\cdot 16 \tau^a X}{2N_f+4N_b}\int \frac{\dd^3 q}{(2\pi)^3}\frac{1-\zeta}{q^3}\rightarrow\frac{2\cdot 16 \tau^a X}{(2N_f+4N_b)\pi^2}(-1+\zeta)\log k\,,
\end{split}
\end{equation}
Evidently, the vertex correction from this diagram is independent of the explicit matrix structure of $X$. For the second diagram, however, we must distinguish between two cases: (a) $X$ couples to the Dirac index via a gamma matrix (i.e., $X\propto c_b\gamma^b$ for coefficients $c_b$ with $b=0,x,y$), and (b) $X$ does not couple to the Dirac indices.

In case (a), we obtain
\begin{equation}\label{eq:yukawa_renormalization_2a}
\begin{split}
    I_{\psi\psi}^{(a)} &= \tau^b\tau^a\tau^b\int \frac{\dd^3 q}{(2\pi)^3}\gamma_\mu G_\psi(q) X G_\psi(q)\gamma_\nu D_{\mu\nu}(-q)\\
    &=-\tau^a\frac{16}{2N_f+4N_b} \int \frac{\dd^3 q}{(2\pi)^3}\gamma_\mu \frac{\slashed{q}X\slashed{q}}{q^5}\gamma_\nu\left(\delta_{\mu\nu}-\zeta\frac{q_\mu q_\nu}{q^2}\right)\\
    &=\tau^a \frac{16}{2N_f+4N_b}\int \frac{\dd^3 q}{(2\pi)^3}\left[\frac{\slashed{q}X\slashed{q}}{q^5}+\zeta \frac{X}{q^3}\right]\\
    &\rightarrow\tau^a X \frac{8}{(2N_f+4N_b)\pi^2}\left(\frac{1}{3}-\zeta\right)\log k\,.
\end{split}
\end{equation}
Conversely, in case (b), the integral evaluates to:
\begin{equation}\label{eq:yukawa_renormalization_2b}
\begin{split}
    I_{\psi\psi}^{(b)} &= \tau^b\tau^a\tau^b\int \frac{\dd^3 q}{(2\pi)^3}\gamma_\mu G_\psi(q) X G_\psi(q)\gamma_\nu D_{\mu\nu}(-q)\\
    &=-\tau^a\frac{16X}{2N_f+4N_b} \int \frac{\dd^3 q}{(2\pi)^3}\gamma_\mu \frac{1}{q^3}\gamma_\nu\left(\delta_{\mu\nu}-\zeta\frac{q_\mu q_\nu}{q^2}\right)\\
    &=-\tau^a\frac{16X}{2N_f+4N_b} \int \frac{\dd^3 q}{(2\pi)^3}\left[ \frac{3-\zeta}{q^3}\right]\\
    &\rightarrow\tau^a X \frac{8}{(2N_f+4N_b)\pi^2}(3-\zeta)\log k\,.
\end{split}
\end{equation}
Therefore, the scaling dimension of the coupling constant $y$ in each scenario is given by:
\begin{equation}
\begin{split}\label{eq:yres}
    (a):\qquad \dim[y^{(a)}] &= \frac{1}{2} - \frac{4}{9N_b\pi^2}+\frac{320}{3(2N_f+4N_b)\pi^2}\,,\\
    (b):\qquad \dim[y^{(b)}] &= \frac{1}{2} - \frac{4}{9N_b\pi^2}+\frac{256}{3(2N_f+4N_b)\pi^2}\,.
\end{split}
\end{equation}
We observe that the fluctuations of the Lagrange multiplier act to reduce the scaling dimension of $y$ (the second term on the right hand side of Eq.~(\ref{eq:yres})), while the gauge field fluctuations act to increase the scaling dimension of $y$ (the third term on the right hand side of Eq.~(\ref{eq:yres})). This is to be contrasted with the result for the critical exponent $\nu$ in Eq.~(\ref{eq:nures}), where both types of fluctuations decrease the value of $\nu$. The large-flavor expansion is not expected to be a reliable estimate of the relative strength of the opposing tendencies in the renormalization of $y$, and so the ultimate fate of $y$ at the $\SO(5)$ fixed point remains unclear. Evaluating Eq.~(\ref{eq:yres}) directly for 
the physically relevant $N_f=2,N_b=2$ yields $\dim[y^{(a)}]\approx 1.378 >0$ and $\dim[y^{(b)}]\approx 1.198$. The observation that the Yukawa coupling is more relevant when coupling to the Dirac indices aligns with the findings in Ref.~\cite{feuerpfeil_2026_higgsa}.

Applying this to the Shastry--Sutherland lattice discussed in \secref{sec:ss_symmetries}, the Yukawa couplings $X^s=(\mu^z\gamma^x,\mu^x\gamma^y)$ for the two Higgs fields $\Phi_{1,2}$---along with the additional symmetry-allowed terms $\mu^x\gamma^x$ and $\mu^z\gamma^y$---all fall under case (a). Consequently, they share the maximum scaling dimension $\dim[y]=\dim[y^{(a)}]\approx 1.378$. These are the most relevant Yukawa couplings, demonstrating that the extra symmetry-allowed terms on the Shastry--Sutherland lattice do not alter the scaling dimension compared to the square lattice. Yukawa couplings with other zeroth-order bilinears, corresponding to $\U(1)$ and $\mathbb{Z}_2$ spin liquids other than the ones considered here, possess either an identical or smaller scaling dimension; therefore, within the large-$N_{f,b}$ framework, they are not expected to change the ultimate fate of the $\SO(5)$ critical point.

The relevance of the Yukawa coupling in the large-$N$ analysis implies that the $\SO(5)$-symmetric fixed point is ultimately unstable in the strict infrared~\cite{Zhou_2024} and the relevant quantum field theory is the one studied in~\cite{Shackleton2021}. However, extensive numerical studies of deconfined criticality have demonstrated that weakly relevant perturbations can generate an extended regime of approximate $\SO(5)$ symmetry and scaling~\cite{Chen_2024,Chester_2024}, giving rise to pseudocritical or ``walking'' behavior~\cite{Nahum2015,Shao2016Science,Nahum2015PRX}. In this regime, renormalization-group flows remain close to the $\SO(5)$-symmetric fixed point over many decades before eventually running away, leading to apparent universal scaling with well-defined effective critical exponents~\cite{Takahashi2024}. From this perspective, the $\SO(5)$ fixed point acts as a controlling multicritical structure rather than a true infrared limit, reconciling the large-$N$ instability with numerical and experimental observations of robust deconfined-like behavior at intermediate scales~\cite{Wang_2017,Hu-2013,Hering-2019,Ferrari-2020,Nomura-2021,LiuESS-2024}.

\section{Conclusion and Discussion}
\label{sec:conclusion}
In this work, we have established a unified fermionic gauge-theory description of deconfined criticality on the square and Shastry--Sutherland lattices, demonstrating that their low-energy continuum theories coincide at the level of the most relevant couplings controlling Higgs transitions out of an $\SU(2)$ $\pi$-flux parent state. Despite the reduced lattice symmetry of the Shastry--Sutherland geometry and the correspondingly enlarged set of symmetry-allowed fermion bilinears, we have shown that all such additional couplings are either irrelevant or share identical scaling dimensions with their square-lattice counterparts within a controlled large-$N$ expansion. As a result, the structure of the Higgs potential, the emergent $\SO(5)$ symmetry unifying N\'eel and valence-bond-solid order, and the leading operator content at criticality are preserved.

We have also proposed a candidate conformal field theory with $\SO(5)$ symmetry: an $\SU(2)$ gauge theory with $N_f=2$ massless, fundamental Dirac fermions, and $N_b = 2$, critical adjoint Higgs scalars. The $N_f=2$, $N_b=0$ case has been extensively studied earlier, with evidence for pseudocritical behavior~\cite{Nahum2015,Wang_2017,WangPRL2021,Zhou_2024,Chen_2024,Meng24,Takahashi2024}. Adding $N_b=2$ adjoint Higgs scalars can reasonably be expected to stabilize the theory to a true conformal theory. This theory is proposed to control the physics near the boundary between the N\'eel and gapless $\mathbb{Z}_2$ spin liquid states.

Our renormalization-group analysis identifies the Yukawa coupling between Dirac spinons and Higgs fields as the leading $\SO(5)$-breaking perturbation, rendering the $\SO(5)$-symmetric fixed point unstable at the lowest energies. This instability may nevertheless be compatible with the growing body of numerical evidence indicating pseudocritical or ``walking'' behavior characterized by approximate $\SO(5)$ symmetry over an extended intermediate regime. Our results suggest that large anomalous dimensions of N\'eel and VBS order parameters may serve as a robust and experimentally relevant diagnostic of fermionic deconfined criticality, even when the ultimate infrared fixed point is unstable. This reinforces the view of the $\SO(5)$ theory as an organizing principle controlling extended pseudocritical regimes in frustrated quantum magnets. Taken together, our results place recent numerical observations of a gapless $\mathbb{Z}_2$ Dirac spin liquid on the Shastry--Sutherland lattice within a coherent and controlled field-theoretic framework, and demonstrate that fermionic deconfined criticality extends beyond the square lattice to frustrated geometries with reduced symmetry.

Several directions emerge naturally from our analysis. First, while the present work establishes the operator content and leading instabilities of the $\SO(5)$ critical theory, a quantitative characterization of the associated pseudocritical regime---including crossover scales, finite-size scaling forms, and multifractal operator spectra---would provide a direct bridge to recent large-scale Monte Carlo studies. In this context, it would be particularly interesting to examine whether lattice-specific perturbations can parametrically enhance the extent of the walking regime on frustrated lattices relative to the square lattice.

Second, our framework can be generalized to incorporate additional symmetry-allowed perturbations such as Dzyaloshinskii--Moriya interactions, further-neighbor exchanges, or explicit breaking of spin-rotation symmetry, which are generically present in material realizations of the Shastry--Sutherland lattice. Understanding how such terms interact with the Higgs mechanism and the emergent gauge structure may shed light on the stability of gapless $\mathbb{Z}_2$ spin liquids in realistic settings.

Third, the field theory developed here provides a natural starting point for studying dynamical probes and experimental signatures of fermionic deconfined criticality, including finite-temperature transport, dynamical structure factors, and responses to pressure or magnetic field in compounds such as SrCu$_2$(BO$_3$)$_2$. More broadly, our results suggest that fermionic deconfined criticality offers a unifying paradigm for organizing quantum phase transitions and proximate spin-liquid physics across a wide class of frustrated quantum magnets, motivating its application to other lattices with reduced symmetry and complex unit cells.

Finally, from a field-theoretic perspective, several extensions of the present analysis are particularly natural. While our conclusions rely on a controlled large-$N$ expansion to determine operator relevance and stability, it would be valuable to complement this approach with alternative analytical techniques that are sensitive to finite-$N$ effects, such as conformal bootstrap methods applied to $\SO(5)$-symmetric or weakly broken fixed points, or functional renormalization-group approaches that can capture runaway flows and walking behavior beyond leading order. Such methods may help quantify the extent to which the pseudocritical regime persists at physical values of $N$ and clarify the nature of the eventual infrared fate.

In addition, the structure of the Majorana--Higgs theory suggests a broader landscape of nearby critical and multicritical points, including the possibility of tuning between distinct Higgs condensates or coupling multiple Higgs sectors. Exploring whether such extensions admit stable fixed points, complex fixed points, or lines of walking behavior may provide further insight into the universality and robustness of fermionic deconfined criticality. More generally, developing a systematic classification of allowed Higgs theories descending from $\SU(2)$ $\pi$-flux states on lattices with reduced symmetry could help identify which aspects of deconfined criticality are truly universal and which depend sensitively on microscopic lattice structure.

\paragraph{Funding information}
This work is supported by the Deutsche Forschungsgemeinschaft (DFG, German Research Foundation) through Project-ID 258499086 -- SFB 1170 and through the W\"urzburg-Dresden Cluster of Excellence on Complexity, Topology and Dynamics in Quantum Matter -- ctd.qmat Project-ID 390858490 -- EXC 2147. The Flatiron Institute is a division of the Simons Foundation. S.S. was supported by the U.S. National Science Foundation grant No. DMR 2245246 and by the Simons Collaboration on Ultra-Quantum Matter, which is a grant from
the Simons Foundation (651440, S.S.). The work of Y.I. was performed in part at the Aspen Center for Physics, which is supported by a grant from the Simons Foundation (1161654, Troyer). This research was also supported in part by grant NSF PHY-2309135 to the Kavli Institute for Theoretical Physics and by the International Centre for Theoretical Sciences (ICTS) for participating in the Discussion Meeting -- Fractionalized Quantum Matter (code: ICTS/DMFQM2025/07). R.T. thanks IIT Madras for a Visiting Faculty Fellow position under the IoE program.  Y.I. acknowledges support from the Abdus Salam International Centre for Theoretical Physics through the Associates Programme, from the Simons Foundation through Grant No.~284558FY19, from IIT Madras through the Institute of Eminence program for establishing QuCenDiEM (Project No. SP22231244CPETWOQCDHOC).

\newpage

\begin{appendix}
\numberwithin{equation}{section}

\section{Projective symmetry group}\label{app:PSGs}
In the following, we present details of the PSGs and \textit{ansätze} for the $\SU(2)$ $\pi$-flux state, the U800 staggered-flux state, and the Z3000 $\mathbb{Z}_2$ state. These are based on the results of~\cite{Maity2024}, but are presented here in a different gauge.
\subsection[\texorpdfstring{${\SU(2)}$}{SU(2)} \texorpdfstring{${\pi}$}{pi}-flux]{\texorpdfstring{$\bm{\SU(2)}$}{SU(2)} \texorpdfstring{$\bm{\pi}$}{pi}-flux}
The PSG is given by
\begin{align}
& U_{T_{2x}}(\vecr,m_x,m_y)=g^{}_{T_{2x}},\;U_{T_{2y}}(\vecr,m_x,m_y)=g^{}_{T_{2y}}\,,\notag\\
& U_{G_x}(\vecr,m_x,m_y)=g^{}_{G_x}\,,\notag\\
& U_{\sigma^{}_{xy}}(\vecr,m_x,m_y)=(-1)^{\delta_{(m_x,m_y),(B,B)}}g^{}_{\sigma^{}_{xy}}\,,\notag\\
& U_{\mathcal{T}}(\vecr,m_x,m_y)=-(-1)^{\delta_{m_x,m_y}}g^{}_{\mathcal{T}}\,,\label{eq:psg_su2_pi}
\end{align}
where $g^{}_{T_{2x}}$, $g^{}_{T_{2y}}$, $g^{}_{G_x}$, $g^{}_{\sigma^{}_{xy}}$ and $g^{}_{\mathcal{T}}$ are global $\SU(2)$ matrices. In this gauge, the structure of the \textit{ansatz} is given by
\begin{equation} \label{eq:SU2_pi-flux}
u_{\vi,\vi+\hatx}=it,\;  u_{\vi,\vi+\haty}=(-1)^{i_x}it .
\end{equation}
This \textit{ansatz} also exists with square lattice symmetry and the associated PSG is given by:
\begin{align}
& U_{T_{x}}(\vi)=(-1)^{i_y}g^{}_{T_x},\;U_{T_{y}}(\vi)=g^{}_{T_y},\notag\\
& U_{P_x}(\vi)=(-1)^{i_x}g^{}_{P_x},\;U_{P_y}(\vi)=(-1)^{i_y}g^{}_{P_y},\notag\\
& U_{R_{\pi/2}}(\vi)=(-1)^{i_x+i_xi_y}g^{}_{R_{\pi/2}},\notag\\
& U_{\mathcal{T}}(\vi)=(-1)^{i_x+i_y}g^{}_{\mathcal{T}}.\label{eq:psg_su2_pi_squre}
\end{align}
\subsection{U800}
The corresponding $\U(1)$ PSG can be obtained from \eqnref{eq:psg_su2_pi} by choosing $g^{}_{T_{2x}}=e^{i\phi^{}_{T_{2x}}\tau^z}$, $g^{}_{T_{2y}}=e^{i\phi^{}_{T_{2y}}\tau^z}$, $g^{}_{G_x}=e^{i\phi^{}_{G_x}\tau^z}i\tau^x$, $g^{}_{\sigma^{}_{xy}}=e^{i\phi^{}_{\sigma_{xy}}\tau^z}i\tau^x$ and $g^{}_{\mathcal{T}}=e^{i\phi^{}_{\mathcal{T}}\tau^z}$ and is given by:
\begin{align}
& U_{T_{2x}}(\vecr,m_x,m_y)=e^{i\phi^{}_{T_{2x}}\tau^z},\;U_{T_{2y}}(\vecr,m_x,m_y)=e^{i\phi^{}_{T_{2y}}\tau^z},\notag\\
& U_{G_x}(\vecr,m_x,m_y)=e^{i\phi^{}_{G_x}\tau^z}i\tau^x,\notag\\
& U_{\sigma^{}_{xy}}(\vecr,m_x,m_y)=(-1)^{\delta_{(m_x,m_y),(B,B)}}e^{i\phi^{}_{\sigma_{xy}}\tau^z}i\tau^x,\notag\\
& U_{\mathcal{T}}(\vecr,m_x,m_y)=-(-1)^{\delta_{m_x,m_y}}e^{i\phi^{}_{\mathcal{T}}\tau^z}.\label{eq:psg_u1}
\end{align}
where $\phi^{}_{T_{2x}},\phi^{}_{T_{2y}},\phi^{}_{G_x},\phi^{}_{\sigma_{xy}},\phi^{}_{\mathcal{T}}\in[0,2\pi)$.  In this gauge, the structure of the \textit{ansatz} is given by:
\begin{align}\label{eq:ansatz_g_2}
    u_{\vi,\vi+\hatx}&=\dot\iota{t}^{}_{s,0}\tau^0-(-1)^{i_x+i_y}{t}^{}_{s,z}\tau^z\notag\,,\\ u_{\vi,\vi+\haty}&=(-1)^{i_x}\dot\iota{t}^{}_{s,0}\tau^0+(-1)^{i_y}{t}^{}_{s,z}\tau^z. 
\end{align}
In square lattice symmetry, the PSG is given by:
\begin{align}
& U_{T_{x}}(\vi)=(-1)^{i_y}e^{i\phi^{}_{T_x}\tau^z}i\tau^x\,,\;U_{T_{y}}(\vi)=e^{i\phi^{}_{T_y}\tau^z}i\tau^x\,,\notag\\
& U_{P_x}(\vi)=(-1)^{i_x}e^{i\phi^{}_{P_x}\tau^z}\,,\;U_{P_y}(\vi)=(-1)^{i_y}e^{i\phi^{}_{P_y}\tau^z}\,,\notag\\
& U_{R_{\pi/2}}(\vi)=(-1)^{i_x+i_xi_y}e^{i\phi^{}_{R_{\pi/2}}\tau^z}i\tau^x\,,\notag\\
& U_{\mathcal{T}}(\vi)=(-1)^{i_x+i_y}e^{i\phi^{}_{\mathcal{T}}\tau^z}\,.\label{eq:psg_u1_squre}
\end{align}

\begin{figure*}	\includegraphics[width=1.0\linewidth]{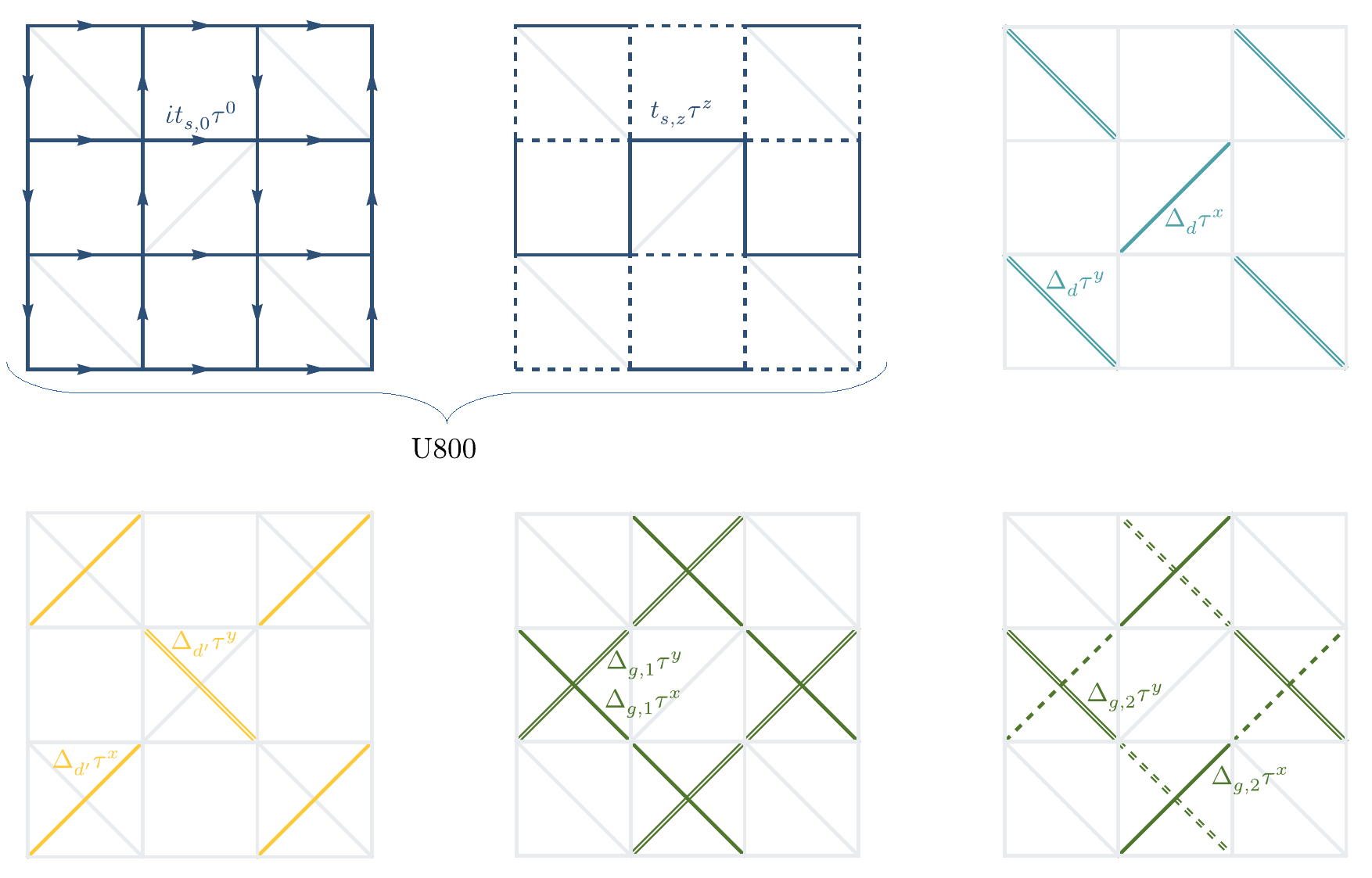}
	\caption{Sign structures of the hopping (also for the $\U(1)$ \textit{ansatz} labeled U800) and pairing amplitudes of the $\mathbb{Z}_2$ \textit{ansatz} labeled by Z3000. For the staggered-flux structure $(\varphi,-\varphi)$, the hoppings are $t^{}_{s,0}=t\sin\left(\frac{\varphi+\pi}{4}\right)$ and $t^{}_{s,z}=t\cos\left(\frac{\varphi+\pi}{4}\right)$. All the dashed links incorporate an additional negative sign.}
	\label{fig:u800_z3000_ansatz}
\end{figure*} 

\begin{figure}  
    \centering	\includegraphics[width=0.75\linewidth]{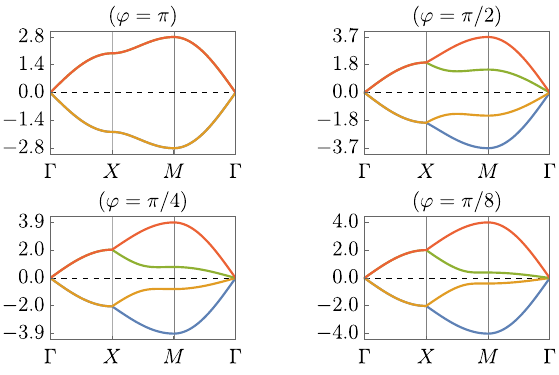}
	\caption{Spinon excitation spectra of U800 for different staggered fluxes $\varphi$ along the high symmetry paths connecting the high-symmetry points given by $\Gamma (0,0)$, X$(\frac{\pi}{2},0)$ and M$(\frac{\pi}{2},\frac{\pi}{2})$. }
	\label{fig:u800_dispersion}
\end{figure} 

The dispersion of U800 for different values of $\varphi$ is shown in \figref{fig:u800_dispersion}, with four Dirac cones at the $\Gamma$ point.

\subsection{Z3000}\label{app:PSGs_Z3000}
The corresponding $\mathbb{Z}_2$ PSG can be obtained from \eqnref{eq:psg_u1} by choosing $\phi^{}_{T_{2x}}=\phi^{}_{T_{2y}}=0$, $\phi^{}_{G_x}=-\pi/4$, $\phi^{}_{\sigma^{}_{xy}}=-\pi/2$, and $\phi^{}_{\mathcal{T}}=\pi/2$ and is given by
\begin{align}
    & U_{T_{2x}}(\vecr,m_x,m_y)=U_{T_{2y}}(\vecr,m_x,m_y)=\tau^0\,,\notag\\
    & U_{G_x}(\vecr,m_x,m_y)=i\frac{1}{\sqrt{2}}(\tau^x+\tau^y)\,,\notag\\
    & U_{\sigma^{}_{xy}}(\vecr,m_x,m_y)=(-1)^{\delta_{(m_x,m_y),(B,B)}}i\tau^y\,,\notag\\
    & U_{\mathcal{T}}(\vecr,m_x,m_y)=-(-1)^{\delta_{m_x,m_y}}i\tau^z\,.\label{eq:psg_z2_oroinal}
\end{align}
In this gauge, the structure of the \textit{ansatz} is given by Eqs.~\eqref{eq:U1_1}--\eqref{eq:Z2_end}. Figure~\ref{fig:u800_z3000_ansatz} illustrates the sign structures of the mean-field hopping and pairing amplitudes of the descendant $\mathbb{Z}_2$ \textit{ansatz} labeled by Z3000 (which is also connected to Z2A$zz$13 on the square lattice). Notice that in the illustrated gauge form, the connection to the parent $\U(1)$ state U800 is explicit as the removal of the pairing terms gives the parent state.

The gaplessness of the Z3000 state is not accidental, but protected by its projective symmetry group (PSG)~\cite{Wen2002,Maity2024}. After the parent $\mathrm{U}(1)$ state is higgsed down to $\mathbb{Z}_2$, the remaining projective implementation of time-reversal and Shastry--Sutherland lattice symmetries forbids any symmetry-preserving fermion bilinear mass term for the Dirac spinons~\cite{Lu-2018}. Consequently, weak perturbations that preserve the full PSG can at most renormalize the Dirac velocities, shift the node positions, or generate higher-derivative corrections, but they cannot open a gap~\cite{Lu-2018}. In this sense, Z3000 is a symmetry-protected gapless $\mathbb{Z}_2$ spin liquid, in direct analogy with the square-lattice Z2A$zz$13 state to which it is connected by PSG and adiabatic interpolation~\cite{Maity2024,Maity2026unified}. More generally, this fits into the symmetry-fractionalization perspective developed for gapless $\mathbb{Z}_2$ spin liquids, where the interplay of time reversal with crystalline symmetries can forbid a fully symmetric gapped spectrum~\cite{Qi-2018,Lu-2018}. Beyond mean field, the $\mathbb{Z}_2$ gauge field is gapped, so the state is not destabilized by gauge confinement; the remaining possible instabilities are therefore associated with sufficiently strong symmetry-breaking perturbations or with a transition out of the Z3000 phase itself.

\section{Perturbations due to pairing terms}\label{app:pairing_terms}
In this Appendix, we present details on the derivation of the perturbations due to the pairing terms $\Delta_{g,1/2}$:
\begin{equation}
\begin{split}
    \delta H =&\sum_{\vecr}\Tr\Big\lbrace(\Delta_{g,1}\tau^y-\Delta_{g,2}\tau^x)[\X_{\vecr,A,0}^\dag+\X_{\vecr,B,1}^\dag][\X_{\vecr+\hatx,B,0}-\X_{\vecr+\hatx,A,1}]+\mathrm{h.c.}\\
    &\qquad +(\Delta_{g,1}\tau^x+\Delta_{g,2}\tau^y)[\X_{\vecr,B,0}^\dag+\X_{\vecr,A,1}^\dag][\X_{\vecr-\hatx,A,0}-\X_{\vecr-\hatx,B,1}]+\mathrm{h.c.}\\
    &\qquad +(\Delta_{g,1}\tau^y+\Delta_{g,2}\tau^x)[\X_{\vecr,B,0}^\dag-\X_{\vecr,A,1}^\dag][\X_{\vecr+\haty,A,0}+\X_{\vecr+\haty,B,1}]+\mathrm{h.c.}\\
    &\qquad +(\Delta_{g,1}\tau^x-\Delta_{g,2}\tau^y)[\X_{\vecr,A,0}^\dag-\X_{\vecr,B,1}^\dag][\X_{\vecr+\haty,B,0}+\X_{\vecr+\haty,A,1}]+\mathrm{h.c.}\Big\rbrace\\
    \approx&\int\dd^2 {\vecr} \Tr\Big\lbrace(\Delta_{g,1}\tau^y-\Delta_{g,2}\tau^x)(\X^\dag \rho^x\mu^z\X-\X^\dag \rho^z\mu^x\X)
    \\
    &\qquad+(\Delta_{g,1}\tau^x+\Delta_{g,2}\tau^y)(\X^\dag \rho^x\mu^z\X+\X^\dag \rho^z\mu^x\X)\\
    &\qquad +(\Delta_{g,1}\tau^y+\Delta_{g,2}\tau^x)(\X^\dag \rho^x\mu^z\X-\X^\dag \rho^z\mu^x\X)\\
    &\qquad +(\Delta_{g,1}\tau^x-\Delta_{g,2}\tau^y)(\X^\dag \rho^x\mu^z\X+\X^\dag \rho^z\mu^x\X)\Big\rbrace\\
    \Rightarrow \delta \mathcal{L}=&2\Delta_{g,1}\Tr[\tau ^x\bar{\X}\left(\gamma^x\mu^z-\gamma^y\mu^x\right)\X]+2\Delta_{g,1}\Tr[\tau^y\bar{\X}\left(\gamma^x\mu^z+\gamma^y\mu^x\right)\X]\,.
\end{split}
\end{equation}
As we can see, to lowest order $\Delta_{g,2}$ vanishes, but there is a non-vanishing gradient term, which we now extract:
\begin{equation}
\begin{split}
    \delta H
    =&\sum_{\vecr}\Delta_{g,2}\Tr\Big\lbrace\tau^x[\X^\dag_{\vecr,A,0}+\X_{\vecr,B,1}^\dag][\X_{\vecr-\haty,B,0}-\X_{\vecr-\haty,A,1}]+[\X^\dag_{\vecr,B,0}-\X_{\vecr,A,1}^\dag][\X_{\vecr+\haty,A,0}+\X_{\vecr+\haty,B,1}]\\
    &\qquad-[\X^\dag_{\vecr,A,0}+\X_{\vecr,B,1}^\dag][\X_{\vecr+\hatx,B,0}-\X_{\vecr+\hatx,A,1}]-[\X^\dag_{\vecr,B,0}-\X_{\vecr,A,1}^\dag][\X_{\vecr-\hatx,A,0}+\X_{\vecr-\hatx,B,1}]\Big\rbrace\\
    &+\Delta_{g,2}\Tr\Big\lbrace\tau^y[\X_{\vecr,B,0}^\dag+\X_{\vecr,A,1}^\dag][\X_{\vecr-\hatx,A,0}-\X_{\vecr-\hatx,B,1}]+[\X_{\vecr,A,0}^\dag-\X_{\vecr,B,1}^\dag][\X_{\vecr+\hatx,B,0}+\X_{\vecr+\hatx,A,1}]\\
    &\qquad -[\X_{\vecr,B,0}^\dag+\X_{\vecr,A,1}^\dag][\X_{\vecr-\haty,A,0}-\X_{\vecr-\haty,B,1}]-[\X_{\vecr,A,0}^\dag-\X_{\vecr,B,1}^\dag][\X_{\vecr+\haty,B,0}+\X_{\vecr+\haty,A,1}]\Big\rbrace\\
    \approx &\Delta_{g,2}\int\dd^2\vecr \Tr\lbrace \tau^x\X^\dag (\mu^y-\rho^y)(i\partial_x+i\partial_y)\X\rbrace + \Tr\lbrace \tau^y\X^\dag (-\mu^y-\rho^y)(i\partial_y-i\partial_x)\X\rbrace\\
    \Rightarrow \delta \mathcal{L}=&\Delta_{g,2}\Tr[\tau^x\bar{\X}(-\gamma^0\mu^y+1)(i\partial_x+i\partial_y)\X+\tau^y\bar{\X}(\gamma^0\mu^y+1)(i\partial_y-i\partial_x)\X]\,.
\end{split}
\end{equation}

\section{Saddle point expansion}\label{app:saddle_point_expansion}
\begin{figure}[t]
    \centering
    \includegraphics[width=0.6\linewidth]{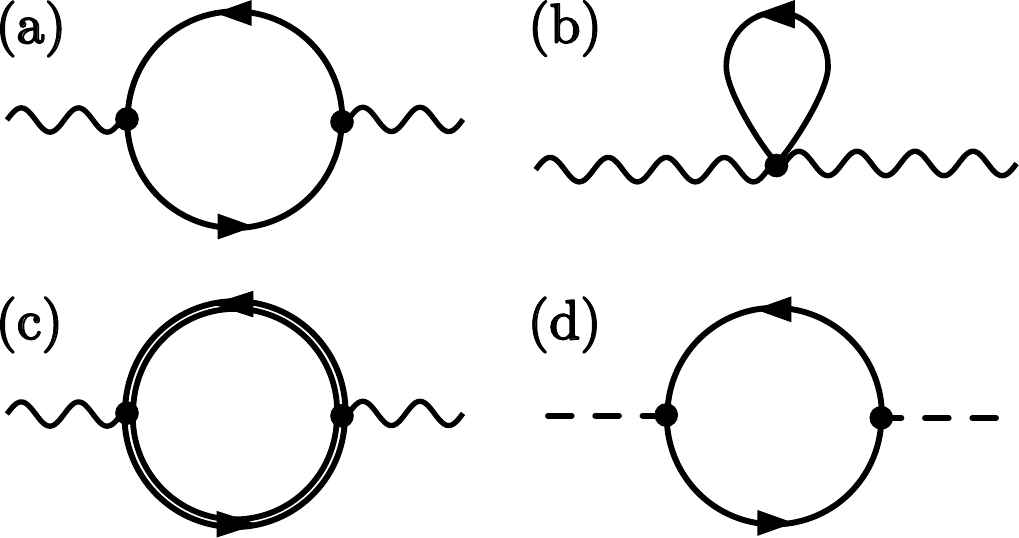}
    \caption{One-loop diagrams that determine the effective action at large $N_f$ and $N_b$. (a), (b) and (c) contribute to the effective action for $A_\mu$, and (d) contributes to the action of the $\lambda$ field.}
    \label{fig:app:saddle_point_expansion_diagrams}
\end{figure}

We present more details on the derivation of the expansion of the free energy around the saddle point up to second order. In the bosonic sector, we obtain the corrections
\begin{equation}
\begin{split}
    &\frac{N_b}{2}\int \frac{\dd^3p}{(2\pi)^3}\frac{\dd^3q}{(2\pi)^3}\Bigg(8G_\Phi(q)\delta_{p,0}\int \frac{\dd^3p'}{(2\pi)^3}A_{\mu}^a(p')A_{\mu}^a(-p')\\
    &\qquad \quad -\frac{1}{2}G_\Phi(q)G_\Phi(q-p)\Big[-3\lambda(p)\lambda(-p)+8(2q-p)_\mu A_\mu^a(p)(2q-p)_\nu A_\nu^a(-p)\Big]\Bigg)
\end{split}
\end{equation}
where the prefactor $N_b/2$ accounts for the $3N_b$ real components of the Higgs fields, and in the fermionic sector
\begin{equation}
    \frac{N_f}{2}\int \frac{\dd^3 p}{(2\pi)^3}\frac{\dd^3 q}{(2\pi)^3}\Tr[G_\psi(q)\gamma^\mu A_\mu^a (p)\tau_a G_{\psi}(p+q)\gamma^\nu A_{\nu}^b(-p)\tau_b]\,.
\end{equation}
The diagrams corresponding to these terms are shown in \figref{fig:app:saddle_point_expansion_diagrams}. These integrals are identical to~\cite{Christos_2024}, and the results are:
\begin{align}
    &\int \frac{\dd^3q}{(2\pi)^3}\frac{1}{q^2+r}=-\frac{\sqrt{r}}{4\pi}\,,\\ 
    &\int \frac{\dd^3 q}{(2\pi)^3}\frac{1}{(q^2+r)((q-p)^2+r)}=\frac{1}{4\pi p}\arctan \frac{p}{2\sqrt{r}}\,,\\
    &\int \frac{\dd^3 q}{(2\pi)^3}\frac{(2q-p)_\mu (2q-p)_\nu}{(q^2+r)((q-p)^2+r)}=-\left(\delta_{\mu\nu}+\frac{p_\mu p_\nu}{p^2}\right)\frac{\sqrt{r}}{4\pi}\nonumber\\
    &\hspace{14.7em}-\left(\delta_{\mu\nu}-\frac{p_\mu p_\nu}{p^2}\right)\left(\frac{4r+p^2}{8\pi p}\arctan\frac{p}{2\sqrt{r}}\right)\,,\\
    &\int \frac{\dd^3q}{(2\pi)^3}\frac{\Tr[\gamma^\mu \slashed{q}\gamma^\nu (\slashed{p}+\slashed{q})]}{q^2(p+q)^2}=\frac{p}{16}\left(\delta_{\mu\nu}-\frac{p_\mu p_\nu}{p^2}\right)\,.
\end{align}
In the fermionic term the trace runs over Dirac and gauge indices; with $\Tr(\tau_a\tau_b)=2\delta_{ab}$ and the last integral above, it contributes $2N_f\,p/16$ to $\Pi_A$, while the bosonic term contributes $4N_b\,p/16$ at $r=0$.

\section{Useful integrals}\label{app:useful_integrals}
In this Appendix, we exemplify the extraction of the logarithmic divergences via dimensional regularization for some of the integrals which appear in our treatment.

\subsection{Fermion self-energy}
We can calculate the fermion self-energy \eqnref{eq:fermion_selfenergy} as 
\begin{equation}
\begin{split}
    \Sigma_\psi(k)&=\tau^a\tau^a\int \frac{\dd^3q}{(2\pi)^3}\gamma_\mu G_\psi(k+q) \gamma_\nu D_{\mu\nu}(-q)\\
    &=3\int \frac{\dd^3q}{(2\pi)^3}\frac{16}{2N_f+4N_b}\gamma_\mu\frac{\slashed{q}+\slashed{k}}{(k+q)^2}\gamma_\nu \frac{1}{q}\left(\delta_{\mu\nu}-\zeta \frac{q_\mu q_\nu}{q^2}\right)\\
    &=3\frac{16}{2N_f+4N_b}\int \frac{\dd^3q}{(2\pi)^3}\left[\gamma_\mu \frac{\slashed{q}+\slashed{k}}{(k+q)^2q}\gamma_\mu- \zeta\frac{\slashed{q}(\slashed{q}+\slashed{k})\slashed{q}}{(k+q)^2q^3}\right]\\
    &=3\frac{16}{2N_f+4N_b}\int \frac{\dd^3q}{(2\pi)^3}\left[ -\frac{\slashed{q}+\slashed{k}}{(k+q)^2q}-\zeta\frac{\slashed{q}(\slashed{q}+\slashed{k})\slashed{q}}{(k+q)^2q^3}\right]\,.
\end{split}
\end{equation}
We split the integral into two parts. We solve each by dimensional regularization by introducing a Feynman parameter $x$ and defining $l=q+(1-x)k$:
\begin{equation}
\begin{split}
    \int \frac{\dd^3 q}{(2\pi)^3}\frac{\slashed{q}+\slashed{k}}{(k+q)^2q}&=\frac{1}{2}\int_0^1 \dd x \int \frac{\dd^3 q}{(2\pi)^3}\frac{x^{-1/2}(\slashed{q}+\slashed{k})}{[(1-x)(k+q)^2+xq^2]^{3/2}}\\
    &=\frac{1}{2}\int_0^1 \dd x \int \frac{\dd^3 l}{(2\pi)^3}\frac{x^{-1/2}(\slashed{l}+x\slashed{k})}{[l^2+x(1-x)k^2]^{3/2}}\\
    &=\int_0^1 \dd x \frac{1}{2(4\pi)^{3/2}}\left[\frac{\sqrt{x}\slashed{k}}{\Gamma(3/2)}\frac{\Gamma(0)}{[x(1-x)k^2]^0}\right]\\
    &\rightarrow-\frac{1}{6\pi^2}\slashed{k}\log k
\end{split}
\end{equation}
In the last line, we performed the dimensional regularization in the following way: Define $\Delta=x(1-x)k^2$ and $\epsilon=\frac{3}{2}-\frac{D}{2}$, then
\begin{equation}
    \frac{\Gamma(\epsilon)}{\Delta^\epsilon}\approx \frac{1/\epsilon}{1+\epsilon \log \Delta}\approx \frac{1}{\epsilon}-\log \Delta \rightarrow -\log k^2 = -2\log k\,.
\end{equation}
The second integral similarly gives
\begin{equation}
\begin{split}
    \int \frac{\dd^3 q}{(2\pi)^3}\frac{\slashed{q}(\slashed{q}+\slashed{k})\slashed{q}}{(k+q)^2q^3}&=\frac{3}{2}\int_0^1 \dd x \int \frac{\dd^3 q}{(2\pi)^3}\frac{\sqrt{x}\slashed{q}(\slashed{q}+\slashed{k})\slashed{q}}{[(1-x)(k+q)^2+xq^2]^{5/2}}\\
    &=\frac{3}{2}\int_0^1 \dd x \sqrt{x}\int \frac{\dd^3 l}{(2\pi)^3}\frac{(\slashed{l}-(1-x)\slashed{k})(\slashed{l}+x\slashed{k})(\slashed{l}-(1-x)\slashed{k})}{[l^2+x(1-x)k^2]^{5/2}}\\
    &\rightarrow \frac{3}{2}\int_0^1 \dd x\frac{\sqrt{x}(-2\log k)}{8\pi\sqrt{\pi}\cdot 2\Gamma(5/2)}\delta_{\mu\nu}\big[-(1-x)(\gamma^\mu \gamma^\nu \slashed{k}+\slashed{k}\gamma^\mu \gamma^\nu )+x\gamma^\mu \slashed{k}\gamma^\nu\big]\\
    &=\int_0^1 \dd x\frac{-2\sqrt{x}\log k}{8\pi^2} \big[-3(1-x)\slashed{k}-x\slashed{k}-3(1-x)\slashed{k}\big]\\
    &=\int_0^1 \dd x \frac{-2\sqrt{x}(5x-6)}{8\pi^2}\slashed{k}\log k\\
    &= \frac{1}{2\pi^2}\slashed{k}\log k\,.
\end{split}
\end{equation}

\subsection{Boson self-energy}

The boson self-energy corrections in \eqnref{eq:I_A1} read 
\begin{equation}
\begin{split}
    I_{A;1}&=8\int \frac{\dd^3 q}{(2\pi)^3}G_\Phi(k+q)D_{\mu\nu}(-q)(2k+q)_\mu (2k+q)_\nu \\
    &= \frac{8\cdot 16}{2N_f+4N_b}\int \frac{\dd^3 q}{(2\pi)^3}\frac{(2k+q)_\mu(2k+q)_\nu}{q(k+q)^2}\left(\delta_{\mu\nu}-\zeta \frac{q_\mu q_\nu}{q^2}\right)\\
    &=\frac{8\cdot 16}{2N_f+4N_b}\int \frac{\dd^3 q}{(2\pi)^3}\left[\frac{(2k+q)^2}{q(k+q)^2}-\zeta \frac{(2k q+q^2)^2}{q^3(k+q)^2}\right]\,.
\end{split}
\end{equation}
We solve both integrals:
\begin{equation}
\begin{split}
    \int \frac{\dd^3 q}{(2\pi)^3}\frac{(2k+q)^2}{q(k+q)^2}&=\frac{1}{2}\int_0^1 \dd x \int \frac{\dd^3 q}{(2\pi)^3}\frac{x^{-1/2}(2k+q)^2}{[(1-x)(k+q)^2+xq^2]^{3/2}}\\ 
    &=\frac{1}{2}\int_0^1 \dd x \int \frac{\dd^3 l}{(2\pi)^3}\frac{x^{-1/2}(l+(1+x)k)^2}{[l^2+x(1-x)k^2]^{3/2}}\\
    &\rightarrow\frac{k^2\log k}{16\pi\sqrt{\pi}\Gamma(3/2)}\int_0^1 \dd x \frac{1}{\sqrt{x}} \left[-2(1+x)^2+3x(1-x)\right]\\
    &=-\frac{10}{12\pi^2}k^2 \log k\,.
\end{split}
\end{equation}
\begin{equation}
\begin{split}
    \int \frac{\dd^3 q}{(2\pi)^3}\frac{(2k q+q^2)^2}{q^3(k+q)^2}&=\frac{3}{2}\int_0^1 \dd x \int \frac{\dd^3 q}{(2\pi)^3}\frac{\sqrt{x}(2kq+q^2)^2}{[(1-x)(k+q)^2+xq^2]^{5/2}}\\
    &=\frac{3}{2}\int_0^1 \dd x \int \frac{\dd^3 l}{(2\pi)^3}\frac{\sqrt{x}[(l+(1+x)k)_\mu(l-(1-x)k)_\mu]^2}{[l^2+x(1-x)k^2]^{5/2}}\\
    &=\frac{3}{2}\int_0^1 \dd x \sqrt{x}\int \frac{\dd^3 l}{(2\pi)^3}\frac{(l^2)^2+4x^2(k\cdot l)^2+(1-x^2)^2k^4-2(1-x^2)l^2k^2}{[l^2+x(1-x)k^2]^{5/2}}\\
    &\rightarrow\frac{3}{2}\frac{k^2\log k}{8\pi\sqrt{\pi}\Gamma(5/2)}\int_0^1 \dd x \sqrt{x} \left[\frac{15x(1-x)}{2}-4x^2+6(1-x^2)\right]\\
    &=\frac{1}{2 \pi^2}k^2\log k\,.
\end{split}
\end{equation}
The Lagrange multiplier fluctuations contribute [\eqnref{eq:I_lambda1}]:
\begin{equation}
\begin{split}
    I_{\lambda;1} &= i^2 \int \frac{\dd^3 q}{(2\pi)^3}G_\Phi(k+q)D_\lambda(-q)\\
    &= -\frac{16}{3N_b}\int \frac{\dd^3 q}{(2\pi)^3}\frac{q^2}{q(k+q)^2}\\
    &= -\frac{8}{3N_b}\int_0^1 \dd x \int \frac{\dd^3 q}{(2\pi)^3}\frac{x^{-1/2}q^2}{[(1-x)(k+q)^2+xq^2]^{3/2}}\\
    &= -\frac{8}{3N_b}\int_0^1 \dd x \int \frac{\dd^3 l}{(2\pi)^3}x^{-1/2}\frac{l^2+(1-x)^2k^2-2(1-x)l\cdot k}{[l^2+x(1-x)k^2]^{3/2}}\\
    &\rightarrow-\frac{8}{3N_b\cdot 8\pi \sqrt{\pi}\Gamma(3/2)}\int_0^1\dd x \frac{1}{\sqrt{x}}\left[\frac{6x(1-x)}{2}-2(1-x)^2\right]k^2\log k\\
    &=\frac{8}{9N_b\pi^2}k^2\log k\,.
\end{split}
\end{equation}

\subsection[\texorpdfstring{${\SO(5)}$}{SO(5)} vertex corrections]{\texorpdfstring{$\bm{\SO(5)}$}{SO(5)} vertex corrections}
The vertex correction due to the gauge field is given by
\begin{equation}
\begin{split}
    I_{A;2}&=3 \int \frac{\dd^3 q}{(2\pi)^3} \gamma_\mu G_\psi(k_1+q) G_\psi(k_2+q) \gamma_\nu D_{\mu\nu}(-q)\\
    &= \frac{3\cdot 16}{2N_f+4N_b}\int \frac{\dd^3 q}{(2\pi)^3}\gamma_\mu \frac{\slashed{k}_1+\slashed{q}}{(k_1+q)^2}\frac{\slashed{k}_2+\slashed{q}}{(k_2+q)^2}\frac{1}{q}\gamma_\nu\left(\delta_{\mu\nu}-\zeta \frac{q_\mu q_\nu}{q^2}\right)\\ 
    &=\frac{3\cdot 16}{2N_f+4N_b}\int \frac{\dd^3 q}{(2\pi)^3}\left[ \frac{-(\slashed{k}_1+\slashed{q})(\slashed{k}_2+\slashed{q})+4(k_1+q)\cdot(k_2+q)}{(k_1+q)^2(k_2+q)^2q}-\zeta \frac{\slashed{q}(\slashed{k}_1+\slashed{q})(\slashed{k}_2+\slashed{q})\slashed{q}}{(k_1+q)^2(k_2+q)^2q^3}\right]\,.
\end{split}
\end{equation}
We will now focus on the case $k_1=k_2\equiv k$. We split the integral into two parts and solve each of them via dimensional regularization:
\begin{equation}
\begin{split}
    I_{A;2,1}&=\int \frac{\dd^3 q}{(2\pi)^3}\frac{3}{(k+q)^2q}\\
    &=\frac{1}{2}\int_0^1 \dd x \int \frac{\dd^3 q}{(2\pi)^3}\frac{3x^{-1/2}}{[(1-x)(k+q)^2+xq^2]^{3/2}}\\
    &=\frac{3}{2}\int_0^1 \dd x \frac{1}{\sqrt{x}}\int \frac{\dd^3 l}{(2\pi)^3}\frac{1}{[l^2+x(1-x)k^2]^{3/2}}\\
    &\rightarrow \frac{3}{2}\int_0^1 \dd x \frac{1}{\sqrt{x}}\frac{1}{8\pi \sqrt{\pi}\Gamma(3/2)}(-2\log k)=-\frac{3}{2\pi^2}\log k\,,
\end{split}
\end{equation}
and 
\begin{equation}
\begin{split}
    I_{A;2,2}&=\int \frac{\dd^3 q}{(2\pi)^3}\frac{q^2}{(k+q)^2q^3}=\frac{1}{3}I_{A;2,1}\,.
\end{split}
\end{equation}
\end{appendix}

\bibliography{qftssmodel.bib}

\end{document}